\documentclass[10pt, conference, compsocconf]{IEEEtran}
\usepackage{subfigure}
\usepackage{graphicx}
\usepackage{amsmath}
\usepackage{bm}
\usepackage{url}
\usepackage{stmaryrd}
\usepackage{balance}
\usepackage{amssymb}
\usepackage{pgfplots}
\usepackage{pifont}
\usepackage[usenames,dvipsnames]{pstricks}
\usepackage{epsfig}
\usepackage{booktabs}
\usepackage{pgffor}
\usepackage{tikz}
\usepackage{multirow}
\usepackage{algorithm}
\usepackage{algorithmic}
\definecolor{purple}{rgb}{0.6,0.2,0.8}

\makeatletter
\newcommand\xleftrightarrow[2][]{%
  \ext@arrow 9999{\longleftrightarrowfill@}{#1}{#2}}
\newcommand\longleftrightarrowfill@{%
  \arrowfill@\leftarrow\relbar\rightarrow}
\makeatother

\newcommand{\mb}{\mathbf}
\newcommand{\mc}{\mathcal}

\newcommand{\ouromt}[0]{\textsc{PNAomg}}
\newcommand{\ourdmt}[0]{\textsc{PNAdmg}}
\newcommand{\ourom}[0]{\textsc{PNAom}}
\newcommand{\ourdm}[0]{\textsc{PNAdm}}
\newcommand{\ourm}[0]{\textsc{Mna}}
\newcommand{\ouro}[0]{\textsc{PNAo}}
\newcommand{\ourd}[0]{\textsc{PNAd}}
\newcommand{\oursup}[0]{\textsc{Mna}\_{no}}
\newcommand{\our}[0]{\textsc{PNA}}

\begin{document}

\title{Partial Network Alignment with Anchor Meta Path and Truncated Generic Stable Matching}


\author{
\IEEEauthorblockN{Jiawei Zhang$^\star$, Weixiang Shao$^\star$, Senzhang Wang$^\dagger$, Xiangnan Kong$^\ast$, Philip S. Yu$^{\star \ddagger}$}
\IEEEauthorblockA{
{$^\star$ University of Illinois at Chicago, Chicago, IL, USA}\\ 
{$^\dagger$ Beihang University, Beijing, China}\\
{$^\ast$ Worcester Polytechnic Institute, Worcester, MA, USA}\\ 
{$^\ddagger$ Institute for Data Science, Tsinghua University, Beijing, China}\\ 
{\{jzhan9, wshao4\}@uic.edu, szwang@cse.buaa.edu.cn, xkong@wpi.edu, psyu@cs.uic.edu}
}
}

%
\maketitle

\begin{abstract}

To enjoy more social network services, users nowadays are usually involved in multiple online social networks simultaneously. The shared users between different networks are called \textit{anchor users}, while the remaining unshared users are named as \textit{non-anchor users}. Connections between accounts of anchor users in different networks are defined as \textit{anchor links} and networks partially aligned by anchor links can be represented as \textit{partially aligned networks}. In this paper, we want to predict anchor links between partially aligned social networks, which is formally defined as the \textit{partial network alignment} problem. The partial network alignment problem is very difficult to solve because of the following two challenges: (1) the lack of \textit{general features} for anchor links, and (2) the ``$one-to-one_{\le}$'' (one to at most one) constraint on anchor links. To address these two challenges, a new method {\our} (\underline{P}artial \underline{N}etwork \underline{A}ligner) is proposed in this paper. {\our} (1) extracts a set of \textit{explicit anchor adjacency features} and \textit{latent topological features} for anchor links based on the \textit{anchor meta path} concept and \textit{tensor decomposition} techniques, and (2) utilizes the \textit{generic stable matching} to identify the \textit{non-anchor users} to prune the redundant anchor links attached to them. Extensive experiments conducted on two real-world partially aligned social networks demonstrate that {\our} can solve the partial network alignment problem very well and outperform all the other comparison methods with significant advantages.

\end{abstract}


\section{Introduction}
\label{sec:intro}

In recent years, online social networks providing various featured services have become an essential part in our lives. To enjoy more social network services, users nowadays are usually involved in multiple online social networks simultaneously \cite{KZY13, ZKY13, ZKY14, ZYZ14} and there can be significant overlaps of users shared by different networks. As pointed out in \cite{DS13}, by the end of 2013, $42\%$ of online adults are using multiple social sites at the same time. For example, $93\%$ of Instagram users are involved in Facebook concurrently and 53\% Twitter users are using Instagram as well \cite{M14}. Formally, the common users involved in different networks simultaneously are named as the ``\textit{anchor users}'' \cite{KZY13}, while the remaining unshared users are called the ``\textit{non-anchor users}'' \cite{ZYZ14}. The connections between accounts of anchor users in different networks are defined as the ``\textit{anchor links}'' \cite{KZY13} and networks partially aligned by anchor links can be represented as ``\textit{partially aligned networks}'' \cite{ZKY14}.

\noindent \textbf{Problem Studied}: In this paper, we want to predict the \textit{anchor links} across \textit{partially aligned networks}, which is formally defined as the ``\textit{partial network alignment}'' problem. 

\textit{Partial network alignment} problem is very important for social networks and can be the prerequisite for many real-world social applications, e.g., link prediction and recommendations \cite{ZKY13, ZKY14, ZYZ14, ZY15-3}, community detection \cite{JZYYL14, ZY15, ZY15-2} and information diffusion \cite{ZZWYX15}. Identifying accounts of anchor users across networks provides the opportunity to compose a more complete social graph with users' information in all the networks they are involved in. Information in the complete social graph is helpful for a better understanding of users' social behavior in online social networks \cite{KZY13, ZY15-2, ZZWYX15}. In addition, via the predicted anchor links, cross-platform information exchange enables new social networks to start their services based on the rich data available in other developed networks. The information transferred from developed networks can help emerging networks \cite{ZKY14, ZY15} to overcome the information shortage problem promisingly \cite{ZKY13, ZKY14, ZY15}.

What's more, the \textit{partial network alignment} problem is a novel problem and different from existing link prediction works, like (1) \textit{traditional intra-network link prediction problems} \cite{SBGAH11, SHAC12}, which mainly focus on predicting links in \textit{one single} social network, (2) \textit{inter-network link transfer problems} \cite{ZKY14}, which can predict links in \textit{one single} network with information from multiple \textit{aligned networks}, and (3) \textit{inferring anchor links across fully aligned networks} \cite{KZY13}, which aims at predicting anchor links across \textit{fully aligned networks}.

The \textit{inferring anchor links across fully aligned networks} problem \cite{KZY13} also studies the \textit{anchor link prediction} problem. However, both the problem setting and method proposed to address the ``\textit{network alignment}'' problem between two \textit{fully aligned networks} in \cite{KZY13} are very ad hoc and have many disadvantages. First of all, the full alignment assumption of social networks proposed in \cite{KZY13} is too strong as fully aligned networks can hardly exist in the real world \cite{ZYZ14}. Secondly, the features extracted for \textit{anchor links} in \cite{KZY13} are proposed for Foursquare and Twitter specifically, which can be hard to get generalized to other networks. Thirdly, the classification based link prediction algorithm used in \cite{KZY13} can suffer from the \textit{class imbalance} problem \cite{LLC10, ME11}. The problem will be more serious when dealing with \textit{partially aligned networks}. Finally, the matching algorithm proposed in \cite{KZY13} is designed specially for \textit{fully aligned networks} and maps all users (including both \textit{anchor} and \textit{non-anchor} users) from one network to another network via the predicted anchor links, which will introduce a large number of non-existing anchor links when applied in the partial network alignment problem.



Totally different from the ``\textit{inferring anchor links across fully aligned networks}'' problem \cite{KZY13}, we study a more general \textit{network alignment} problem in this paper. Firstly, networks studied in this paper are partially aligned \cite{ZYZ14}, which contain large number of anchor and non-anchor users \cite{ZYZ14} at the same time. Secondly, networks studied are not confined to Foursquare and Twitter social networks. A minor revision of the ``\textit{partial network alignment}'' problem can be mapped to many other existing tough problems, e.g., large biology network alignment \cite{BGGSW09}, entity resolution in database integration \cite{BG07}, ontology matching \cite{ES07}, and various types of entity matching in online social networks \cite{PFRE13}. Thirdly, the \textit{class imbalance} problem will be addressed via link sampling effectively in the paper. Finally, the constraint on \textit{anchor links} is ``$one-to-one_{\le}$'' (i.e., each user in one network can be mapped to at most one user in another network). Across partially aligned networks, only anchor users can be connected by anchor links. Identifying the non-anchor users from networks and pruning all the predicted potential anchor links connected to them is a novel yet challenging problem. The ``$one-to-one_{\le}$'' constraint on anchor links can distinguish the ``\textit{partial network alignment}'' problem from most existing link prediction problems. For example, in traditional link prediction and link transfer problems \cite{SBGAH11, SHAC12, ZKY14}, the constraint on links is ``\textit{many-to-many}'', while in the ``\textit{anchor link inference}'' problem \cite{KZY13} across fully aligned networks, the constraint on \textit{anchor links} is strict ``\textit{one-to-one}''.

To solve the ``\textit{partial network alignment}'' problem, a new method, {\our} (\underline{P}artial \underline{N}etwork \underline{A}ligner), is proposed in this paper. {\our} exploits the concept of \textit{anchor meta paths} \cite{ZYZ14, SBGAH11} and utilizes the \textit{tensor decomposition} \cite{MJE12, KB09} technique to obtain a set of \textit{explicit anchor adjacency features} and \textit{latent topological features}. In addition, {\our} generalizes the traditional \textit{stable matching} to support partially aligned network through \textit{self-matching} and \textit{partial stable matching} and introduces the a novel matching method, \textit{generic stable matching}, in this paper.

The rest of this paper is organized as follows. In Section~\ref{sec:formulation}, we will give the definition of some important concepts and formulate the \textit{partial network alignment} problem. {\our} method will be introduces in Sections~\ref{sec:feature_extraction}-\ref{sec:truncation}. Section~\ref{sec:experiment} is about the experiments. Related works will be given in Section~\ref{sec:relatedworks}. Finally, we conclude the paper in Section~\ref{sec:conclusion}.

\section{Problem Formulation} 
\label{sec:formulation}

Before introducing the method {\our}, we will first define some important concepts and formulate the \textit{partial network alignment} problem in this section.

\subsection{Terminology Definition}

\noindent \textbf{Definition 1} (Heterogeneous Social Networks): A \textit{heterogeneous social network} can be represented as $G = (\mathcal{V}, \mathcal{E})$, where $\mathcal{V} = \bigcup_i \mathcal{V}_i$ contains the sets about various kinds of nodes, while $\mathcal{E} = \bigcup_j \mathcal{E}_j$ is the set of different types of links among nodes in $\mathcal{V}$.

\noindent \textbf{Definition 2} (Aligned Heterogeneous Social Networks): Social networks that share common users are defined as the \textit{aligned heterogeneous social networks}, which can be represented as $\mathcal{G} = (G_{set}, A_{set})$, where $G_{set} = (G^{(1)}, G^{(2)}, \cdots, G^{(n)})$ is the set of $n$ different \textit{heterogeneous social networks} and $A_{set} = (\mathcal{A}^{(1,2)}, \mathcal{A}^{(1,3)}, \cdots, \mathcal{A}^{((n-1), n)})$ is the sets of undirected \textit{anchor links} between networks in $G_{set}$.

\noindent \textbf{Definition 3} (Anchor Link): Given two social networks $G^{(i)}$ and $G^{(j)}$, link $(u^{(i)}, v^{(j)})$ is an \textit{anchor link} between $G^{(i)}$ and $G^{(j)}$ iff ($u^{(i)} \in \mathcal{U}^{(i)}$) $\land$ ($v^{(j)} \in \mathcal{U}^{(j)}$) $\land$ ($u^{(i)}$ and $v^{(j)}$ are accounts of the same user), where $\mathcal{U}^{(i)}$ and $\mathcal{U}^{(j)}$ are the user sets of $G^{(i)}$ and $G^{(j)}$ respectively.

\noindent \textbf{Definition 4} (Anchor Users and Non-anchor Users): Users who are involved in two social networks, e.g., $G^{(i)}$ and $G^{(j)}$, simultaneously are defined as the \textit{anchor users} between $G^{(i)}$ and $G^{(j)}$. \textit{Anchor users} in $G^{(i)}$ between $G^{(i)}$ and $G^{(j)}$ can be represented as $\mathcal{U}^{(i)}_{\mathcal{A}^{(i,j)}} = \{u^{(i)} | u^{(i)} \in \mathcal{U}^{(i)}, \exists v^{(j)} \in \mathcal{U}^{(j)}, \mbox{ and }\\(u^{(i)}, v^{(j)}) \in \mathcal{A}^{(i,j)}\}$. Meanwhile, the \textit{non-anchor user} in $G^{(i)}$ between $G^{(i)}$ and $G^{(j)}$ are those who are involved in $G^{(i)}$ only and can be represented as $\mathcal{U}^{(i)}_{-\mathcal{A}^{(i,j)}} = \mathcal{U}^{(i)} - \mathcal{U}^{(i)}_{\mathcal{A}^{(i,j)}}$. Similarly, the \textit{anchor users} and \textit{non-anchor users} in $G^{(j)}$ between $G^{(j)}$ and $G^{(i)}$ can be defined as $\mathcal{U}^{(j)}_{\mathcal{A}^{(i,j)}}$ and $\mathcal{U}^{(j)}_{-\mathcal{A}^{(i,j)}}$ respectively.

\noindent \textbf{Definition 5} (Full Alignment, Partial Alignment and Isolated): Given two social networks $G^{(i)}$ and $G^{(j)}$, if users in both $G^{(i)}$ and $G^{(j)}$ are all anchor users, i.e., $\mathcal{U}^{(i)} = \mathcal{U}^{(i)}_{\mathcal{A}^{(i,j)}}$ and $\mathcal{U}^{(j)} = \mathcal{U}^{(j)}_{\mathcal{A}^{(i,j)}}$, then $G^{(i)}$ and $G^{(j)}$ are \textit{fully aligned}; if users in both of these two networks are all non-anchor users, i.e., $\mathcal{U}^{(i)} = \mathcal{U}^{(i)}_{-\mathcal{A}^{(i,j)}}$ and $\mathcal{U}^{(j)} = \mathcal{U}^{(j)}_{-\mathcal{A}^{(i,j)}}$, then these two networks are \textit{isolated}; otherwise, they are \textit{partially aligned}.

\noindent \textbf{Definition 6} (Bridge Nodes): Besides users, many other kinds of nodes can be shared between different networks, which are defined as the \textit{bridge nodes} in this paper. The \textit{bridge nodes} shared between $G^{(i)}$ and $G^{(j)}$ can be represented as $\mathcal{B}^{(i,j)} = \{v | (v \in (\mathcal{V}^{(i)} - \mathcal{U}^{(i)})) \land (v \in (\mathcal{V}^{(j)} - \mathcal{U}^{(j)}))\}$.

The social networks studied in this paper can be any \textit{partially aligned social networks} and we use Foursquare, Twitter as a example to illustrate the studied problem and the proposed method. Users in both Foursquare and Twitter can make friends with other users, write posts, which can contain words, timestamps, and location checkins. In addition, users in Foursquare can also create lists of locations that they have visited/want to visit in the future. As a result, Foursquare and Twitter can be represented as \textit{heterogeneous social network} $G = (\mathcal{V}, \mathcal{E})$. In Twitter $\mathcal{V} = \mathcal{U} \cup \mathcal{P} \cup \mathcal{W} \cup \mathcal{T} \cup \mathcal{L}$ and in Foursquare $\mathcal{V} = \mathcal{U} \cup \mathcal{P} \cup \mathcal{W} \cup \mathcal{T} \cup \mathcal{I} \cup \mathcal{L}$, where $\mathcal{U}$, $\mathcal{P}$, $\mathcal{W}$, $\mathcal{T}$, $\mathcal{I}$ and $\mathcal{L}$ are the nodes of users, posts, words, timestamps, lists and locations. While in Twitter, the heterogeneous link set $\mathcal{E} = \mathcal{E}_{u,u} \cup \mathcal{E}_{u,p} \cup \mathcal{E}_{p,w} \cup \mathcal{E}_{p,t} \cup \mathcal{E}_{p,l}$ and in Foursquare $\mathcal{E} = \mathcal{E}_{u,u} \cup \mathcal{E}_{u,p} \cup \mathcal{E}_{p,w} \cup \mathcal{E}_{p,t} \cup \mathcal{E}_{p,l} \cup \mathcal{E}_{u,i} \cup \mathcal{E}_{i,l}$. The \textit{bridge nodes} shared between Foursquare and Twitter include the \textit{common locations}, \textit{common words} and \textit{common timestamps}.


\subsection{Problem Statement}

\noindent \textbf{Definition 7} (Partial Network Alignment): For any two given \textit{partially aligned heterogeneous social networks}, e.g., $\mathcal{G} = ((G^{(i)}, G^{(j)}), (\mathcal{A}^{(i,j)}))$, part of the known \textit{anchor links} between $G^{(i)}$ and $G^{(j)}$ are represented as $\mathcal{A}^{(i,j)}$. Let $\mathcal{U}^{(i)}$, $\mathcal{U}^{(j)}$ be the user sets of $G^{(i)}$ and $G^{(j)}$ respectively, the set of other potential \textit{anchor links} between $G^{(i)}$ and $G^{(j)}$ can be represented as $\mathcal{L}^{(i,j)} = \{(u^{(i)}, v^{(j)}) | (u^{(i)} \in \mathcal{U}^{(i)}) \land (v^{(j)} \in \mathcal{U}^{(j)}\} - \mathcal{A}^{(i,j)}$. We solve the \textit{partial network alignment} problem as a \textit{link classification} problem, where existing and non-existing \textit{anchor links} are labeled as ``+1'' and ``-1'' respectively. In this paper, we aim at building a model $\mathcal{M}$ with the existing \textit{anchor links} $\mathcal{A}^{(i,j)}$, which will be applied to predict potential \textit{anchor links} in $\mathcal{L}^{(i,j)}$. In model $\mathcal{M}$, we want to determine both \textit{labels} and \textit{existence probabilities} of \textit{anchor links} in $\mathcal{L}^{(i,j)}$.



%

\section{Feature Extraction and Anchor Link Prediction} \label{sec:feature_extraction}
\textit{Supervised link prediction} method has been widely used in research due to its excellent performance and the profound supervised learning theoretical basis. In supervised link prediction, links are labeled differently according to their physical meanings, e.g., existing vs non-existent \cite{ZYZ14}, friends vs enemies \cite{WS12}, trust vs distrust \cite{YTYXL13}, positive attitude vs negative attitude \cite{YCZC13}. With information in the networks, a set of heterogeneous features can be extracted for links in the training set, which together with the labels are used to build the link prediction model $\mathcal{M}$. 



In this section, we will introduce different categories of general features extracted for \textit{anchor links} across \textit{partially aligned networks}, which include a set of \textit{explicit anchor adjacency features} based on \textit{anchor meta paths} and the ``\textit{latent topological feature vector}'' extracted via \textit{anchor adjacency tensor} decomposition.



\subsection{Traditional Intra-Network Meta Path}

Traditional \textit{meta paths} are mainly defined based on the \textit{social network schema} of one single network \cite{SBGAH11, SHYYW11}.

\noindent \textbf{Definition 8} (Social Network Schema): For a given network $G$, its \textit{schema} is defined as $S_G = (\mathcal{T}_G, \mathcal{R}_G)$, where $\mathcal{T}_G$ and $\mathcal{R}_G$ are the sets of node types and link types in $G$ respectively.

\noindent \textbf{Definition 9} (Meta Path): Based on the schema of network $G$, i.e., $S_G = (\mathcal{T}_G, \mathcal{R}_G)$, the traditional intra-network \textit{meta path} in $G$ is defined as $\Phi = T_1 \xrightarrow{R_1} T_2 \xrightarrow{R_2} \cdots \xrightarrow{R_{k-1}} T_k$, where $T_i \in \mathcal{T}_G, i \in \{1, 2, \cdots, k\}$ and $R_j \in \mathcal{R}_G, j \in \{1, 2, \cdots, k-1\}$ \cite{SBGAH11, SHYYW11}.

For example, according to the networks introduced in Section~\ref{sec:formulation}, we can define the network schema of Twitter as $S_G = (\{User, Post, Word, Timestamp, List, Location\}, \{Follow, \\ Write, Create, Contain, At, Checkin\})$. Based on the schema, ``\textit{User - Location - User}'' is a \textit{meta path} of length 2 connecting user nodes in the network via location node and path ``\textit{Alice} - \textit{San Jose} - \textit{Bob}'' is an instance of such meta path in the network, where \textit{Alice}, \textit{Bob} and \textit{San Jose} are the users and location in the network.



\begin{figure}[t]
\centering
    \begin{minipage}[l]{0.8\columnwidth}
      \centering
      \includegraphics[width=\textwidth]{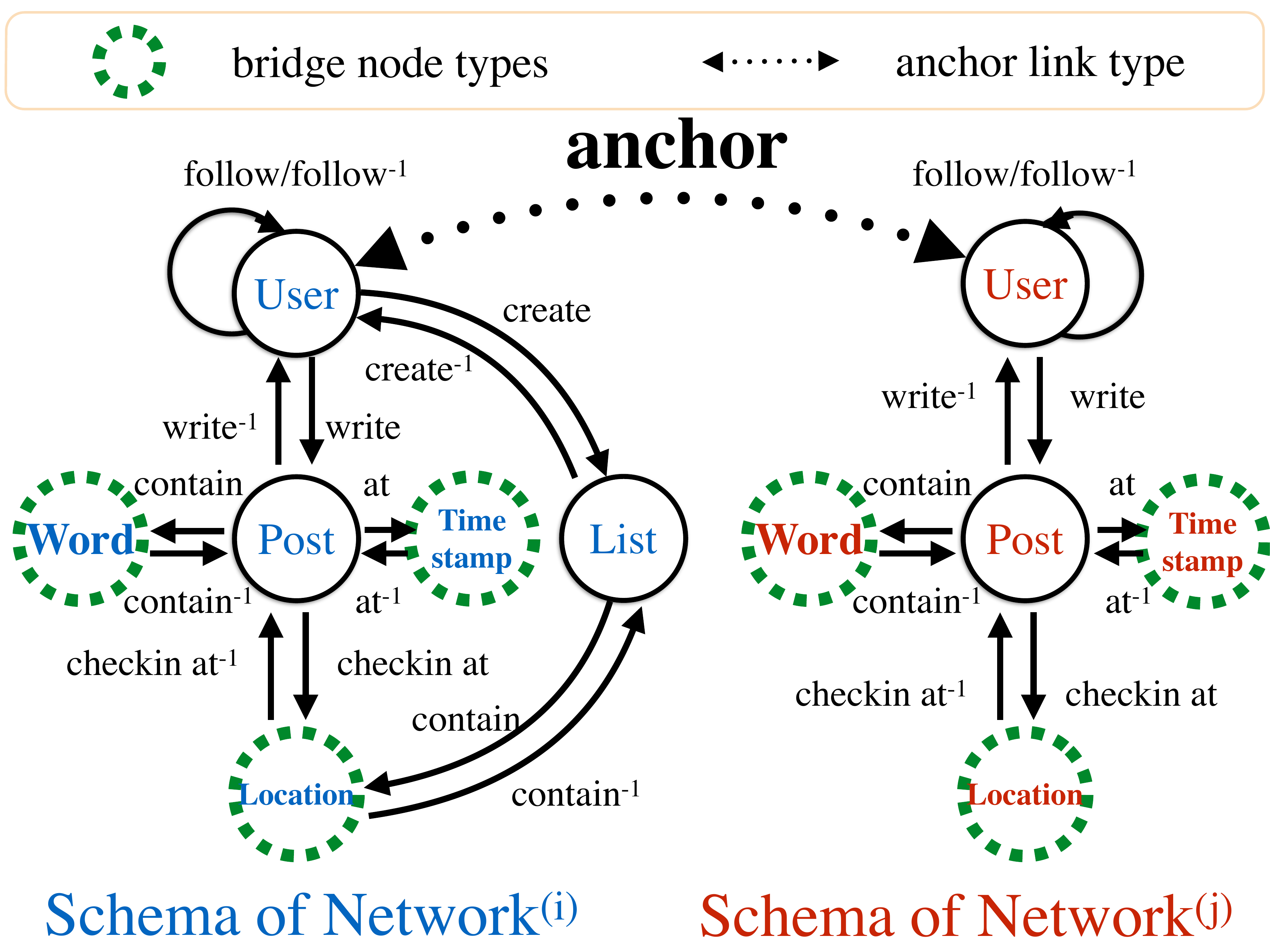}
    \end{minipage}
  \caption{Schema of aligned heterogeneous network.}\label{fig:schema}
\end{figure}

\subsection{Inter-Network Anchor Meta Path}
Traditional Intra-network \textit{meta paths} defined based on one single network cannot be applied to address the inter-network \textit{partial network alignment} problem directly. To overcome such a problem, in this subsection, we will define the concept of \textit{anchor meta paths} and introduce a set of \textit{inter-network anchor meta paths} \cite{ZYZ14} across partially aligned networks.


\noindent \textbf{Definition 10} (Aligned Social Network Schema): Given the \textit{partially aligned networks}: $\mathcal{G} = (G_{set}, A_{set})$, let $S_{G^{(i)}} = (\mathcal{T}_{G^{(i)}}, \mathcal{R}_{G^{(i)}})$ be the \textit{schema} of network $G^{(i)} \in G_{set}$, the \textit{schema} of \textit{partially aligned networks} $\mathcal{G}$ can be defined as $S_{\mathcal{G}} = \left(\bigcup_i \mathcal{T}_{G^{(i)}}, (\bigcup_i \mathcal{R}_{G^{(i)}}) \cup \{Anchor\}\right)$, where $\{Anchor\}$ is the \textit{anchor link} type. 

An example of the schema about two \textit{partially aligned social networks}, e.g., $G^{(i)}$ (e.g., Foursquare) and $G^{(j)}$ (e.g., Twitter), is shown in Figure~\ref{fig:schema}, where the schema of these two aligned networks are connected by the anchor link type and the green dashed circles are the shared \textit{bridge nodes} between $G^{(i)}$ and $G^{(j)}$.



\noindent \textbf{Definition 11} (AMP: Anchor Meta Path): Based on the \textit{aligned social network schema}, \textit{anchor meta paths} connecting users across $\mathcal{G}$ is defined to be $\Psi = T_1 \xrightarrow{R_1} T_2 \xrightarrow{R_2} \cdots \xrightarrow{R_{k-1}} T_k$, where $T_1$ and $T_{k}$ are the ``User'' node type in two \textit{partially aligned social networks} respectively. To differentiate the \textit{anchor link} type from other link types in the \textit{anchor meta path}, the direction of $R_i$ in $\Psi$ will be bidirectional if $R_i = Anchor, i \in \{1, 2, \cdots, k-1\}$, i.e., $T_i \xleftrightarrow{R_i} T_j$.

Via the instances of \textit{anchor meta paths}, users across \textit{aligned social networks} can be extensively connected to each other. In the two partially aligned social networks (e.g., $\mathcal{G} = ((G^{(i)}, G^{(j)}), (\mathcal{A}^{(i,j)}))$) studied in this paper, various \textit{anchor meta paths} from $G^{(i)}$ (i.e., Foursquare) and $G^{(j)}$ (i.e., Twitter) can be defined as follows:

\begin{itemize}
\item \textit{Common Out Neighbor Anchor Meta Path} ($\Psi_1$): $User^{(i)}$ $\xrightarrow{follow}$  $User^{(i)}$ $\xleftrightarrow{Anchor}$ $User^{(j)}$ $\xleftarrow{follow}$  $User^{(j)}$ or ``$\mathcal{U}^{(i)} \to \mathcal{U}^{(i)} \leftrightarrow \mathcal{U}^{(j)} \gets \mathcal{U}^{(j)}$'' for short.

\item \textit{Common In Neighbor Anchor Meta Path} ($\Psi_2$): $User^{(i)}$ $\xleftarrow{follow}$  $User^{(i)}$ $\xleftrightarrow{Anchor}$ $User^{(j)}$ $\xrightarrow{follow}$  $User^{(j)}$ or ``$\mathcal{U}^{(i)} \gets \mathcal{U}^{(i)} \leftrightarrow \mathcal{U}^{(j)} \to \mathcal{U}^{(j)}$'' .

\item \textit{Common Out In Neighbor Anchor Meta Path} ($\Psi_3$): $User^{(i)}$ $\xrightarrow{follow}$  $User^{(i)}$ $\xleftrightarrow{Anchor}$ $User^{(j)}$ $\xrightarrow{follow}$  $User^{(j)}$ or ``$\mathcal{U}^{(i)} \to \mathcal{U}^{(i)} \leftrightarrow \mathcal{U}^{(j)} \to \mathcal{U}^{(j)}$''.

\item \textit{Common In Out Neighbor Anchor Meta Path} ($\Psi_4$): $User^{(i)}$ $\xleftarrow{follow}$  $User^{(i)}$ $\xleftrightarrow{Anchor}$ $User^{(j)}$ $\xleftarrow{follow}$  $User^{(j)}$ or ``$\mathcal{U}^{(i)} \gets \mathcal{U}^{(i)} \leftrightarrow \mathcal{U}^{(j)} \gets \mathcal{U}^{(j)}$''.

\end{itemize}

These above \textit{anchor meta paths} are all defined based the ``User'' node type only across \textit{partially aligned social networks}. Furthermore, there can exist many other \textit{anchor meta paths} consisting of user node type and other \textit{bridge node} types from Foursquare to Twitter, e.g., Location, Word and Timestamp.

\begin{itemize}

\item \textit{Common Location Checkin Anchor Meta Path 1} ($\Psi_5$): $User^{(i)}$ $\xrightarrow{write}$  $Post^{(i)}$ $\xrightarrow{checkin\ at}$  $Location$ $\xleftarrow{checkin\ at}$ $Post^{(j)}$ $\xleftarrow{write}$ $User^{(j)}$ or ``$\mathcal{U}^{(i)} \to \mathcal{P}^{(i)} \to \mathcal{L} \gets\mathcal{P}^{(j)} \gets \mathcal{U}^{(j)}$''.

\item \textit{Common Location Checkin Anchor Meta Path 2} ($\Psi_6$): $User^{(i)}$ $\xrightarrow{create}$  $List^{(i)}$ $\xrightarrow{contain}$  $Location$ $\xleftarrow{checkin\ at}$ $Post^{(j)}$ $\xleftarrow{write}$ $User^{(j)}$ or ``$\mathcal{U}^{(i)} \to \mathcal{I}^{(i)} \to \mathcal{L} \gets\mathcal{P}^{(j)} \gets \mathcal{U}^{(j)}$''.

\item \textit{Common Timestamps Anchor Meta Path} ($\Psi_7$): $User^{(i)}$ $\xrightarrow{write}$  $Post^{(i)}$ $\xrightarrow{at}$  $Time$ $\xleftarrow{at}$ $Post^{(j)}$ $\xleftarrow{write}$ $User^{(j)}$ or ``$\mathcal{U}^{(i)} \to \mathcal{P}^{(i)} \to \mathcal{T} \gets\mathcal{P}^{(j)} \gets \mathcal{U}^{(j)}$''.

\item \textit{Common Word Usage Anchor Meta Path} ($\Psi_8$): $User^{(i)}$ $\xrightarrow{write}$  $Post^{(i)}$ $\xrightarrow{contain}$  $Word$ $\xleftarrow{contain}$ $Post^{(j)}$ $\xleftarrow{write}$ $User^{(j)}$ or ``$\mathcal{U}^{(i)} \to \mathcal{P}^{(i)} \to \mathcal{W} \gets\mathcal{P}^{(j)} \gets \mathcal{U}^{(j)}$''.

\end{itemize}

\subsection{Explicit Anchor Adjacency Features}
Based on the above defined \textit{anchor meta paths}, different kinds of anchor meta path based adjacency relationship can be extracted from the network. In this paper, we define the new concepts of \textit{anchor adjacency score}, \textit{anchor adjacency tensor} and \textit{explicit anchor adjacency features} to describe such relationships among users across \textit{partially aligned social networks}.

\noindent \textbf{Definition 12} (Anchor Meta Path Instance): Based on \textit{anchor meta path} $\Psi = T_1 \xrightarrow{R_1} T_2 \xrightarrow{R_2} \cdots \xrightarrow{R_{k-1}} T_k$, path $\psi = n_1-n_2-\cdots-n_{k-1}-n_k$ is an instance of $\Psi$ iff $n_i$ is an instance of node type $T_i$, $i \in \{1, 2, \cdots, k\}$ and $(n_i, n_{i+1})$ is an instance of link type $R_i$, $\forall i \in \{1, 2, \cdots, k - 1\}$.

\noindent \textbf{Definition 13} (AAS: Anchor Adjacency Score): The \textit{anchor adjacency} score is quantified as the number of \textit{anchor meta path instances} of various \textit{anchor meta paths} connecting users across networks. The \textit{anchor adjacency score} between $u^{(i)} \in \mathcal{U}^{(i)}$ and $v^{(j)} \in \mathcal{U}^{(j)}$ based on meta path $\Psi$ is defined as:
$$AAS_{\Psi}(u^{(i)}, v^{(j)}) = \left| \{ \psi | (\psi \in \Psi) \land (u^{(i)} \in T_1) \land (v^{(j)} \in T_k) \} \right|,$$
where path $\psi$ starts and ends with node types $T_1$ and $T_k$ respectively and $\psi \in \Psi$ denotes that $\psi$ is a path instance of meta path $\Psi$.


The \textit{anchor adjacency scores} among all users across \textit{partially aligned networks} can be stored in the \textit{anchor adjacency matrix} as follows.

\noindent \textbf{Definition 14} (AAM: Anchor Adjacency Matrix): Given a certain \textit{anchor meta path}, $\Psi$, the \textit{anchor adjacency matrix} between $G^{i}$ and $G^{j}$ can be defined as $\mb{A}_{\Psi} \in \mathbb{N}^{|\mathcal{U}^{(i)}| \times |\mathcal{U}^{(j)}|}$ and $A(l, m) = AAS_{\Psi}(u_l^{(i)}, u^{(j)}_m), u_l^{(i)} \in \mathcal{U}^{(i)}, u^{(j)}_m \in \mathcal{U}^{(j)}.$

Multiple \textit{anchor adjacency matrix} can be grouped together to form a \textit{high-order tensor}. A \textit{tensor} is a multidimensional array and an N-order \textit{tensor} is an element of the tensor product of $N$ vector spaces, each of which can have its own coordinate system. As a result, an $1$-order \textit{tensor} is a vector, a $2$-order \textit{tensor} is a matrix and \textit{tensors} of three or higher order are called the \textit{higher-order tensor} \cite{KB09, MJE12}.

\noindent \textbf{Definition 15} (AAT: Anchor Adjacency Tensor): Based on meta paths in $\{\Psi_1, \Psi_2, \cdots, \Psi_8\}$, we can obtain a set of \textit{anchor adjacency matrices} between users in two \textit{partially aligned networks} to be $\{\mb{A}_{\Psi_1}, \mb{A}_{\Psi_2}, \cdots, \mb{A}_{\Psi_8}\}$. With $\{\mb{A}_{\Psi_1},$ $\mb{A}_{\Psi_2},$ $\cdots,$ $\mb{A}_{\Psi_8}\}$, we can construct a 3-order \textit{anchor adjacency tensor} $\mathcal{X} \in \mathbb{R}^{|\mathcal{U}^{(i)}| \times |\mathcal{U}^{(j)}| \times 8}$, where the $i_{th}$ layer of $\mathcal{X}$ is the \textit{anchor adjacency matrix} based on \textit{anchor meta path} $\Psi_i$, i.e., $\mathcal{X}(:,:,i) = \mb{A}_{\Psi_i}, i \in \{1, 2, \cdots, 8\}$.

Based on the \textit{anchor adjacency tensor}, a set of \textit{explicit anchor adjacency features} can be extracted for \textit{anchor links} across \textit{partially aligned social networks}.

\noindent \textbf{Definition 16} (EAAF: Explicit Anchor Adjacency Features): For a certain \textit{anchor link} $(u^{(i)}_l, u^{(j)}_m)$, the \textit{explicit anchor adjacency feature vectors} extracted based on the \textit{anchor adjacency tensor} $\mathcal{X}$ can be represented as $\mb{x} = [x_1, x_2, \cdots, x_8]$ (i.e., the \textit{anchor adjacency scores} between $u^{(i)}_l$ and $u^{(j)}_m$ based on $8$ different \textit{anchor meta paths}), where $x_k = \mathcal{X}(l, m, k), k \in \{1, 2, \cdots, 8\}.$

%
%

\subsection{Latent Topological Feature Vectors Extraction}

\textit{Explicit anchor adjacency features} can express manifest properties of the connections across \textit{partially aligned networks} and are the \textit{explicit topological features}. Besides explicit topological connections, there can also exist some hidden common connection patterns \cite{YCZC13} across \textit{partially aligned networks}. In this paper, we also propose to extract the \textit{latent topological feature vectors} from the \textit{anchor adjacency tensor}.


As proposed in \cite{KB09, MJE12}, a \textit{higher-order tensor} can be decomposed into a \textit{core tensor}, e.g., $\mathcal{G}$, multiplied by a matrix along each mode, e.g., $\mb{A}, \mb{B}, \cdots, \mb{Z}$, with various \textit{tensor decomposition methods}, e.g., Tucker decomposition \cite{KB09}. For example, the \textit{3-order anchor adjacency tensor} $\mathcal{X}$ can be decomposed into three matrices $\mb{A} \in \mathbb{R}^{\mathcal{U}^{(i)} \times P}$, $\mb{B} \in \mathbb{R}^{\mathcal{U}^{(j)} \times Q}$ and $\mb{C} \in \mathbb{R}^{8 \times R}$ and a core tensor $\mathcal{G} \in \mathbb{R}^{P \times Q \times R}$, where $P, Q, R$ are the number of columns of matrices $\mb{A}, \mb{B}, \mb{C}$ \cite{KB09}: 
\begin{align*}
\mathcal{X} &= \sum_{p=1}^P \sum_{q=1}^Q \sum_{r=1}^R g_{pqr} \mb{a}_p \circ \mb{b}_q \circ \mb{c}_r = [\mathcal{G}; \mb{A}, \mb{B}, \mb{C}],
\end{align*}
where $\mb{a}_p \circ \mb{b}_q$ denotes the vector outer product of $\mb{a}_p$ and $\mb{b}_q$.

Each row of $\mb{A}$ and $\mb{B}$ represents a \textit{latent topological feature vector} of users in $\mathcal{U}^{(i)}$ and $\mathcal{U}^{(j)}$ respectively  \cite{MJE12}. Method HOSVD introduced in \cite{KB09} is applied to achieve these decomposed matrices in this paper.

\begin{figure}[t]
\centering
    \begin{minipage}[l]{0.7\columnwidth}
      \centering
      \includegraphics[width=\textwidth]{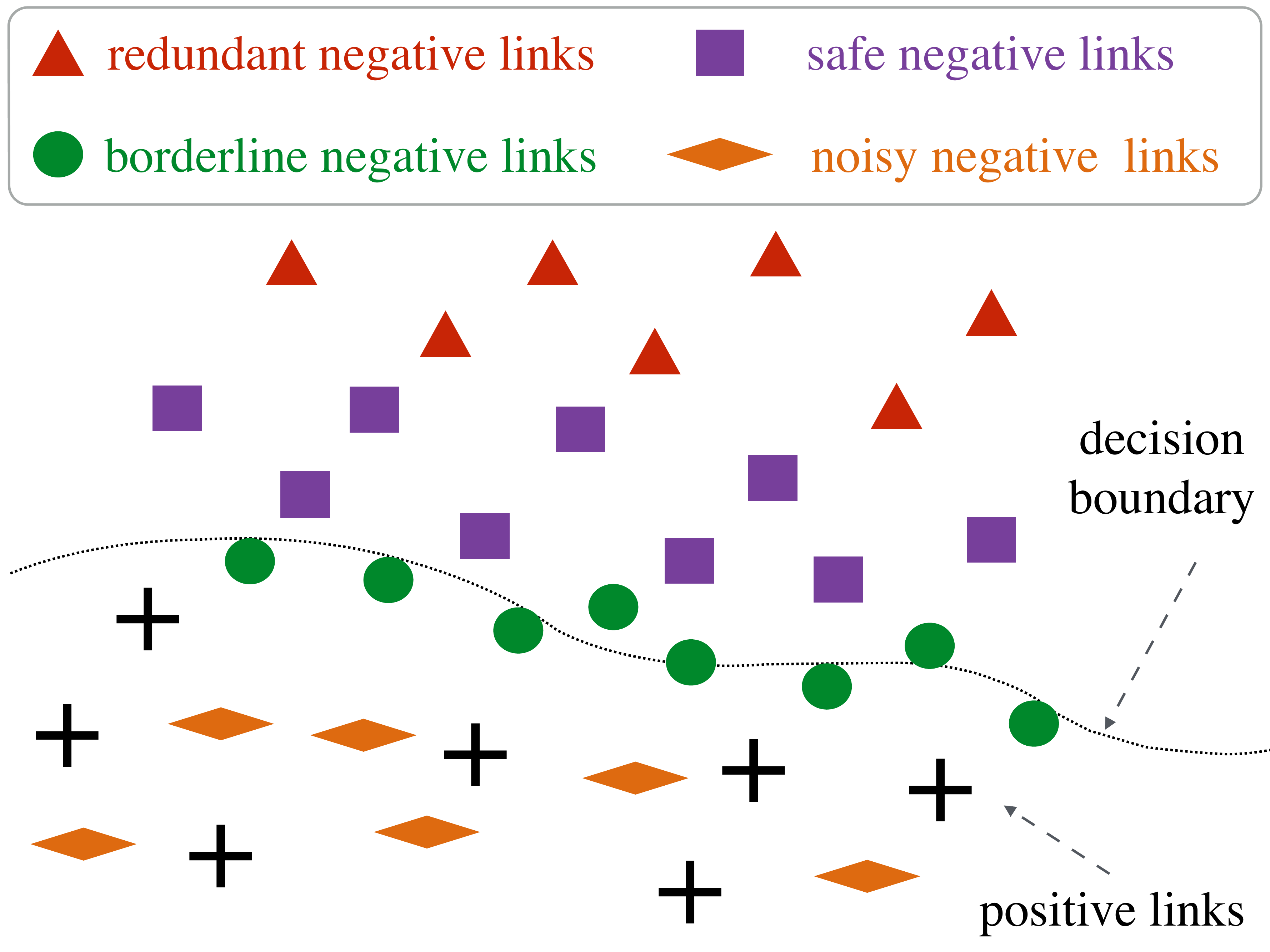}
    \end{minipage}
  \caption{Instance distribution in feature space.}\label{fig:under_and_over_sampling_2}
\end{figure}

\subsection{Class Imbalance Link Prediction}

Based on the extracted features, various supervised link prediction models \cite{KZY13, ZKY14, ZYZ14} can be applied to infer the potential anchor links across networks. As proposed in \cite{ME11, LLC10}, conventional supervised link prediction methods \cite{TGHL13}, can suffer from the \textit{class imbalance} problem a lot. To address the problem, two effective methods (\textit{down sampling} \cite{KM97} and \textit{over sampling} \cite{CBHK02}) are applied.

\textit{Down sampling} methods aim at deleting the \textit{unreliable negative instances} from the training set. In Figure~\ref{fig:under_and_over_sampling_2}, we show the distributions of training links in the feature space, where negative links can be divided into $4$ different categories \cite{KM97}: (1) \textit{noisy links}: links mixed in the positive links; (2) \textit{borderline links}: links close to the decision boundary; (3) \textit{redundant links}: links which are too far away from the decision boundary in the negative region; and (4) \textit{safe links}: links which are helpful for determining the classification boundary.



Different heuristics have been proposed to remove the \textit{noisy instances} and \textit{borderline instances}, which are detrimental for the learning algorithms. In this paper, we will use the method called \textit{Tomek links} proposed in \cite{T76, KM97}. For any two given instances $\mb{x}_1$ and $\mb{x}_2$ of different labels, pair $(\mb{x}_1, \mb{x}_2)$ is called a \textit{tomek link} if there exists no other instances, e.g., $\mb{z}$, such that $d(\mb{x}_1, \mb{z}) < d(\mb{x}_1, \mb{x}_2)$ and $d(\mb{x}_2, \mb{z}) < d(\mb{x}_1, \mb{x}_2)$. Examples that participate in \textit{Tomek links} are either borderline or noisy instances \cite{T76, KM97}. As to the \textit{redundant instances}, they will not harm correct classifications as their existence will not change the classification boundary but they can lead to extra classification costs. To remove the \textit{redundant instances}, we propose to create a \textit{consistent subset} $\mathcal{C}$ of the training set, e.g., $\mathcal{S}$ \cite{KM97}. Subset $\mathcal{C}$ is \textit{consistent} with $\mathcal{S}$ if classifiers built with $\mathcal{C}$ can correctly classify instances in $\mathcal{S}$. Initially, $\mathcal{C}$ consists of all positive instances and one randomly selected negative instances. A classifier, e.g., $kNN$, built with $\mathcal{C}$ is applied to $\mathcal{S}$, where instances that are misclassified are added into $\mathcal{C}$. The final set $\mathcal{C}$ contains the \textit{safe links}.

Another method to overcome the \textit{class imbalance} problem is to \textit{over sample} the \textit{minority class}. Many \textit{over sampling} methods have been proposed, e.g., \textit{over sampling with replacement}, \textit{over sampling with ``synthetic'' instances} \cite{CBHK02}: the minority class is over sampled by introducing new ``synthetic'' examples along the line segment joining $m$ of the $k$ nearest minority class neighbors for each minority class instances. Parameter $k$ is usually set as $5$, while the value of $m$ can be determined according to the ratio to \textit{over sample} the minority class. For example, if the minority class need to be \textit{over sampled} $200\%$, then $m=2$. The instance to be created between a certain example $\mb{x}$ and one of its nearest neighbor $\mb{y}$ can be denoted as $\mb{x} + \mb{\theta}^T(\mb{x} - \mb{y})$, where $\mb{x}$ and $\mb{y}$ are the feature vectors of two instances and $\mb{\theta}^T$ is the transpose of a coefficient vector containing random numbers in range $[0, 1]$.

\section{Anchor Link Pruning with Generic Stable Matching} \label{sec:truncation}

In this section, we will introduce the anchor link pruning methods in details, which include (1) candidate pre-pruning, (2) brief introduction to the traditional stable matching, and (3) the \textit{generic stable matching} method proposed in this paper, which generalizes the concept of traditional stable matching through both \textit{self matching} and \textit{partial stable matching}.

\subsection{Candidate Pre-Pruning}

Across two \textit{partially aligned social networks}, users in a certain network can have a large number of potential \textit{anchor link candidates} in the other network, which can lead to great time and space costs in predicting the anchor links. The problem can be even worse when the networks are of large scales, e.g., containing million even billion users, which can make the \textit{partial network alignment} problem unsolvable. To shrink size of the candidate set, we propose to conduct \textit{candidate pre-pruning} of links in the test set with users' profile information (e.g., names and hometown).

\begin{figure}[t]
\centering
    \begin{minipage}[l]{1.0\columnwidth}
      \centering
      \includegraphics[width=\textwidth]{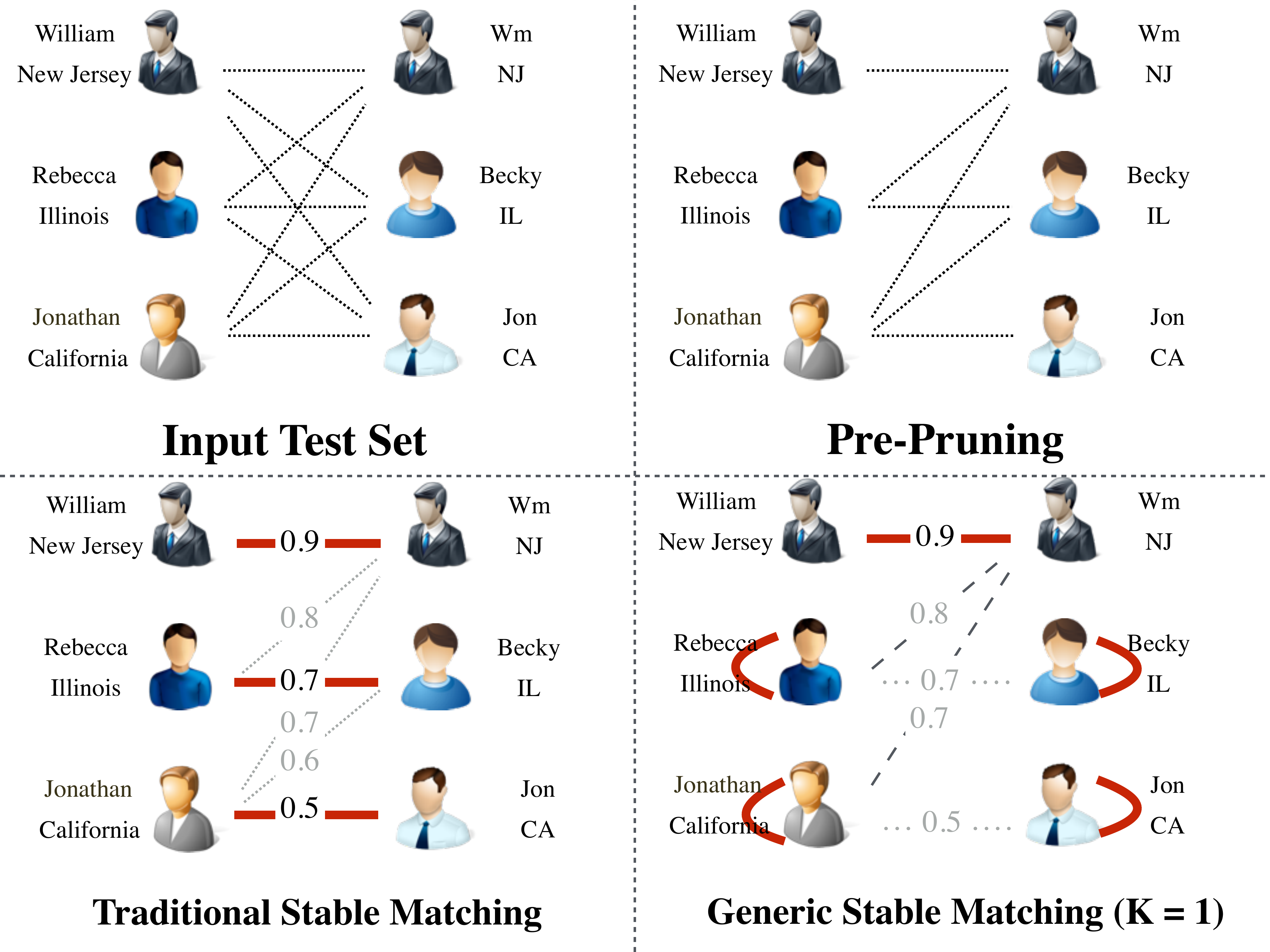}
    \end{minipage}
  \caption{Partial network alignment with pruning.}\label{fig:nash_equilibrium}
\end{figure}

 As shown in Figure~\ref{fig:nash_equilibrium}, in the given input test set, users are extensively connected with all their potential partners in other networks via anchor links. For each users, we propose to prune their potential candidates according to the following heuristics:


\begin{itemize}
\item \textit{profile pre-pruning}: users' profile information shared across \textit{partially aligned social networks}, e.g., Foursquare and Twitter, can include username and hometown \cite{ZL13}. Given an anchor link $(u^{(i)}_l, u^{(j)}_m) \in \mathcal{L}$, if the username and hometown of $u^{(i)}_l$ and $u^{(j)}_m$ are totally different, e.g., cosine similarity scores are $0$, then link $(u^{(i)}_l, u^{(j)}_m)$ will be pruned from test set $\mathcal{L}$.



\item \textit{EAAF pruning}: based on the \textit{explicit anchor adjacency tensor} $\mathcal{X}$ extracted in Section~\ref{sec:feature_extraction}, for a given link $(u^{(i)}_l, u^{(j)}_m) \in \mathcal{L}$, if its extracted \textit{explicit anchor adjacency features} are all $0$, i.e., $\mathcal{X}(l, m, x) = 0, x \in \{1, 2, \cdots, 8\}$, then link $(u^{(i)}_l, u^{(j)}_m)$ will be pruned from test set $\mathcal{L}$.

\end{itemize}



\begin{figure}[t]
\centering
    \begin{minipage}[l]{0.8\columnwidth}
      \centering
      \includegraphics[width=\textwidth]{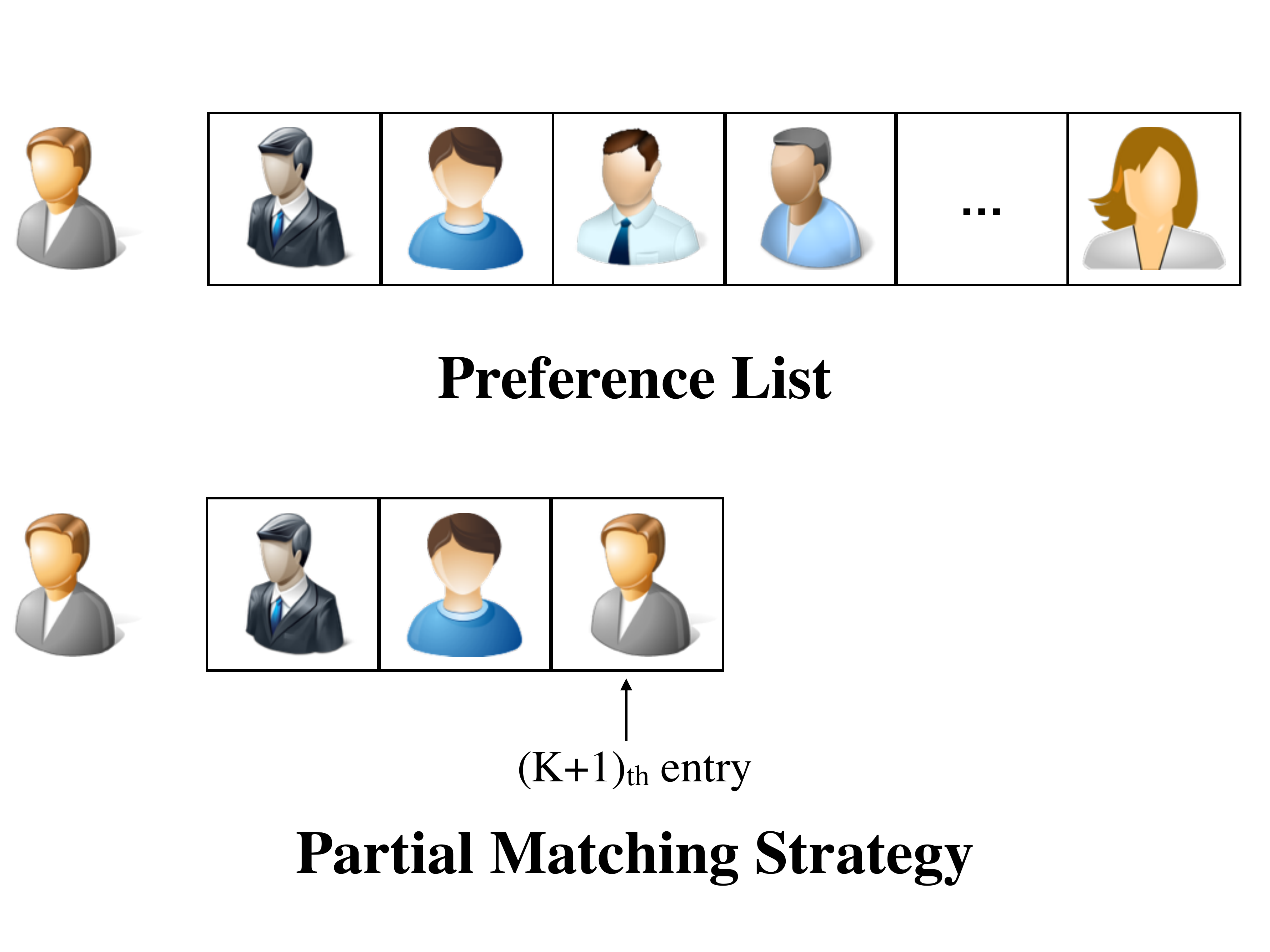}
    \end{minipage}
  \caption{Example of partial matching strategy(K=2).}\label{fig:truncated_strategy}
\end{figure}

\subsection{Traditional Stable Matching}

Meanwhile, as proposed in \cite{KZY13}, the \textit{one-to-one} constraint of anchor links across \textit{fully aligned social networks} can be met by pruning extra potential \textit{anchor link candidates} with \textit{traditional stable matching}. In this subsection, we will introduce the concept of traditional \textit{stable matching} briefly.





Given the user sets $\mathcal{U}^{(1)}$ and $\mathcal{U}^{(2)}$ of two \textit{partially aligned social networks} $G^{(1)}$ and $G^{(2)}$, each user in $\mathcal{U}^{(1)}$(or $\mathcal{U}^{(2)}$) has his preference over users in $\mathcal{U}^{(2)}$(or $\mathcal{U}^{(1)}$). Term $v_j P^{(1)}_{u_i} v_k$ is used to denote that $u_i \in \mathcal{U}^{(1)}$ prefers $v_j$ to $v_k$ for simplicity, where $v_j, v_k \in \mathcal{U}^{(2)}$ and $P^{(1)}_{u_i}$ is the preference operator of ${u_i} \in \mathcal{U}^{(1)}$. Similarly, we can use term $u_i P^{(2)}_{v_j} u_k$ to denote that $v_j \in \mathcal{U}^{(2)}$ prefers $u_i$ to $u_k$ in $\mathcal{U}^{(1)}$ as well.



\noindent \textbf{Definition 17} (Matching): Mapping $\mu: \mathcal{U}^{(1)} \cup \mathcal{U}^{(2)} \to \mathcal{U}^{(1)} \cup \mathcal{U}^{(2)}$ is defined to be a \textit{matching} iff (1) $|\mu(u_i)| = 1, \forall u_i \in \mathcal{U}^{(1)}$ and $\mu(u_i) \in \mathcal{U}^{(2)}$; (2) $|\mu(v_j)| = 1, \forall v_j \in \mathcal{U}^{(2)}$ and $\mu(v_j) \in \mathcal{U}^{(1)}$; (3) $\mu(u_i) = v_j$ iff $\mu(v_j) = u_i$.

\noindent \textbf{Definition 18} (Blocking Pair): A pair $(u_i, v_j)$ is a a \textit{blocking pair} of matching $\mu$ if $u_i$ and $v_j$ prefers each other to their mapped partner, i.e., $(\mu(u_i) \neq v_j) \land (\mu(v_j) \neq u_i)$ and $(v_j P^{(1)}_{u_i} \mu(u_i)) \land (u_i P^{(2)}_{v_j} \mu(v_j))$.


\noindent \textbf{Definition 19} (Stable Matching): Given a matching $\mu$, $\mu$ is \textit{stable} if there is no  \textit{blocking pair} in the matching results \cite{DF81}.

As introduced in \cite{KZY13}, the \textit{stable matching} can be obtained with the Gale-Shapley algorithm proposed in \cite{GS62}.

\subsection{Generic Stable Matching}

Stable matching based method proposed in \cite{KZY13} can only work well in \textit{fully aligned social networks}. However, in the real world, few social networks are fully aligned and lots of users in social networks are involved in one network only, i.e., \textit{non-anchor users}, and they should not be connected by any anchor links. However, traditional \textit{stable matching} method cannot identify these \textit{non-anchor users} and remove the predicted \textit{potential anchor links} connected with them. To overcome such a problem, we will introduce the \textit{generic stable matching} to identify the \textit{non-anchor users} and prune the anchor link results to meet the $one-to-one_{\le}$ constraint.

In {\our}, we introduce a novel concept, \textit{self matching}, which allows users to be mapped to themselves if they are discovered to be \textit{non-anchor users}. In other words, we will identify the \textit{non-anchor users} as those who are mapped to themselves in the final matching results.

\noindent \textbf{Definition 20} (Self Matching): For the given two partially aligned networks $G^{(1)}$ and $G^{(2)}$, user $u_i \in \mathcal{U}^{(1)}$, can have his preference $P^{(1)}_{u_i}$ over users in $\mathcal{U}^{(2)} \cup \{u_i\}$ and $u_i$ preferring $u_i$ himself denotes that $u_i$ is an \textit{non-anchor user} and prefers to stay unconnected, which is formally defined as \textit{self matching}.


Users in one social network will be matched with either partners in other social networks or themselves according to their preference lists (i.e., from high preference scores to low preference scores). Only partners that users prefer over themselves will be \textit{accepted} finally, otherwise users will be matched with themselves instead.

\noindent \textbf{Definition 21} (Acceptable Partner): For a given \textit{matching} $\mu: \mathcal{U}^{(1)} \cup \mathcal{U}^{(2)} \to \mathcal{U}^{(1)} \cup \mathcal{U}^{(2)}$, the mapped partner of users $u_i \in \mathcal{U}^{(1)}$, i.e., $\mu(u_i)$, is \textit{acceptable} to $u_i$ iff $\mu(u_i)P^{(1)}_{u_i}u_i$.

To cut off the partners with very low \textit{preference scores}, we propose the \textit{partial matching strategy} to obtain the promising partners, who will participate in the matching finally.

\noindent \textbf{Definition 22} (Partial Matching Strategy): The \textit{partial matching strategy} of user $u_i \in \mathcal{U}^{(1)}$, i.e., $Q^{(1)}_{u_i}$, consists of the first $K$ the \textit{acceptable partners} in $u_i$'s preference list $P^{(1)}_{u_i}$, which are in the same order as those in $P^{(1)}_{u_i}$, and $u_i$ in the $(K+1)_{th}$ entry of $Q^{(1)}_{u_i}$. Parameter $K$ is called the \textit{partial matching rate} in this paper.

An example is given in Figure~\ref{fig:truncated_strategy}, where to get the top $2$ promising partners for the user, we place the user himself at the $3_{rd}$ cell in the preference list. All the remaining potential partners will be cut off and only the top $3$ users will participate in the final matching.



Based on the concepts of \textit{self matching} and \textit{partial matching strategy}, we define the concepts of \textit{partial stable matching} and \textit{generic stable matching} as follow.

\noindent \textbf{Definition 23} (Partial Stable Matching): For a given \textit{matching} $\mu$, $\mu$ is (1) \textit{rational} if $\mu(u_i) Q^{(1)}_{u_i} u_i, \forall u_i \in \mathcal{U}^{(1)}$ and $\mu(v_j) Q^{(2)}_{v_j} v_j, \forall v_j \in \mathcal{U}^{(2)}$, (2) \textit{pairwise stable} if there exist no \textit{blocking pairs} in the matching results, and (3) \textit{stable} if it is both \textit{rational} and \textit{pairwise stable}.

\noindent \textbf{Definition 24} (Generic Stable Matching): For a given \textit{matching} $\mu$, $\mu$ is a \textit{generic stable matching} iff $\mu$ is a \textit{self matching} or $\mu$ is a \textit{partial stable matching}.

As example of \textit{generic stable matching} is shown in the bottom two plots of Figure~\ref{fig:nash_equilibrium}. \textit{Traditional stable matching} can prune most non-existing anchor links and make sure the results can meet \textit{one-to-one} constraint. However, it preserves the anchor links (Rebecca, Becky) and (Jonathan, Jon), which are connecting \textit{non-anchor users}. In \textit{generic stable matching} with parameter $K=1$, users will be either connected with their most preferred partner or stay \textit{unconnected}. Users ``William'' and ``Wm'' are matched as link (William, Wm) has the highest score. ``Rebecca'' and ``Jonathan'' will prefer to stay \textit{unconnected} as their most preferred partner ``Wm'' is connected with ``William'' already. Furthermore, ``Becky'' and ``Jon'' will stay \textit{unconnected} as their most preferred partner ``Rebecca'' and ``Jonathan'' prefer to stay \textit{unconnected}. In this way, \textit{generic stable matching} can further prune the non-existing anchor links (Rebecca, Becky) and (Jonathan, Jon).


The \textit{truncated generic stable matching} results can be achieved with the \textit{Generic Gale-Shapley} algorithm as given in Algorithm~\ref{alg:gale_shapley}.
%

\begin{algorithm}[t]
\caption{Generic Gale-Shapley Algorithm}
\label{alg:gale_shapley}
\begin{algorithmic}[1]
	\REQUIRE		user sets of aligned networks: $\mathcal{U}^{(1)}$ and $\mathcal{U}^{(2)}$. \\
\qquad  classification results of potential anchor links in $\mathcal{L}$\\
\qquad  known anchor links in $\mathcal{A}^{(1,2)}$\\
\qquad  truncation rate $K$\\
\ENSURE a set of inferred anchor links $\mc{L}'$ \\

\STATE	Initialize the preference lists of users in $\mathcal{U}^{(1)}$ and $\mathcal{U}^{(2)}$ with predicted existence probabilities of links in $\mathcal{L}$ and known anchor links in $\mathcal{A}^{(1,2)}$, whose existence probabilities are $1.0$
\STATE	construct the truncated strategies from the preference lists
\STATE	Initialize all users in $\mathcal{U}^{(1)}$ and $\mathcal{U}^{(2)}$ as \textit{free}
\STATE	$\mc{L}' = \emptyset$
\WHILE{$\exists$ \textit{free} $u^{(1)}_i$ in $\mathcal{U}^{(1)}$ and  $u^{(1)}_i$'s truncated strategy is non-empty}
\STATE	Remove the top-ranked account $u^{(2)}_j$ from  $u^{(1)}_i$'s truncated strategy
\IF	{$u^{(2)}_j$==$u^{(1)}_i$}
\STATE	$\mc{L}' = \mc{L}' \cup \{ (u^{(1)}_i, u^{(1)}_i)\}$
\STATE	Set $u^{(1)}_i$ as \textit{stay unconnected}
\ELSE
\IF{ $u^{(2)}_j$ is \textit{free}} 
\STATE	$\mc{L}' = \mc{L}' \cup \{ (u^{(1)}_i, u^{(2)}_j)\}$
\STATE	Set $u^{(1)}_i$ and $u^{(2)}_j$ as \textit{occupied}
\ELSE
\STATE	$\exists u^{(1)}_p$ that $u^{(2)}_j$ is occupied with.
\IF{ $u^{(2)}_j$ prefers $u^{(1)}_i$ to $u^{(1)}_p$}
\STATE $\mc{L}' = (\mc{L}' - \{ (u^{(1)}_p,u^{(2)}_j) \}) \cup \{ (u^{(1)}_i,u^{(2)}_j) \}$
\STATE Set $u^{(1)}_p$ as \textit{free} and $u^{(1)}_i$ as \textit{occupied}
\ENDIF
\ENDIF
\ENDIF
\ENDWHILE
\end{algorithmic}
\end{algorithm}
\section{Experiments}\label{sec:experiment}

\begin{table}[t]
\caption{Properties of the Heterogeneous Networks}
\label{tab:dataset}
\centering
\begin{tabular}{clrr}
\toprule
&&\multicolumn{2}{c}{network}\\
\cmidrule{3-4}
&property &\textbf{Twitter}	&\textbf{Foursquare}	 \\
\midrule 
\multirow{3}{*}{\# node}
&user		& 5,223	& 5,392 \\
&tweet/tip	& 9,490,707	& 48,756 \\
&location	& 297,182	& 38,921 \\
\midrule 
\multirow{3}{*}{\# link}
&friend/follow		&164,920	&76,972 \\
&write		& 9,490,707	& 48,756 \\
&locate		& 615,515	& 48,756 \\
\bottomrule
\end{tabular}
\end{table}
%

To demonstrate the effectiveness of {\our} in predicting \textit{anchor links} for partially aligned heterogeneous social networks, we conduct extensive experiments on two real-world heterogeneous social networks: Foursquare and Twitter. This section includes three parts: (1) dataset description, (2) experiment settings, and (3) experiment results.

\subsection{Dataset Description}
The datasets used in this paper include: Foursquare and Twitter, which were crawled during November 2012 \cite{KZY13, ZKY13, ZKY14, ZYZ14}. More detailed information about these two datasets is shown in Table~\ref{tab:dataset} and in \cite{KZY13, ZKY13, ZKY14, ZYZ14}. The number of anchor links crawled between Foursquare and Twitter is $3,388$ and $62.83\%$ Foursquare users are \textit{anchor users}.
%
%
%
%


\begin{figure*}[t]
\centering
\subfigure[AUC: alignment rate]{ \label{eg_fig_6_1}
    \begin{minipage}[l]{0.8\columnwidth}
      \centering
      \includegraphics[width=1.0\textwidth]{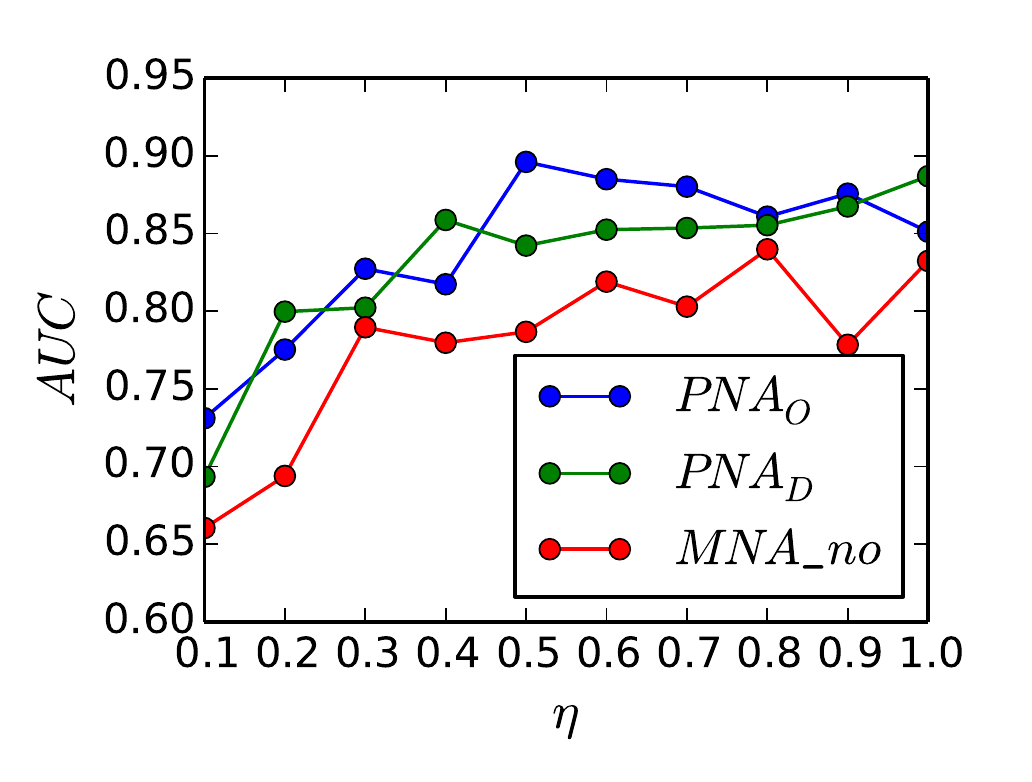}
    \end{minipage}
}
\subfigure[AUC: negative positive rate]{ \label{eg_fig_6_2}
    \begin{minipage}[l]{0.8\columnwidth}
      \centering
      \includegraphics[width=1.0\textwidth]{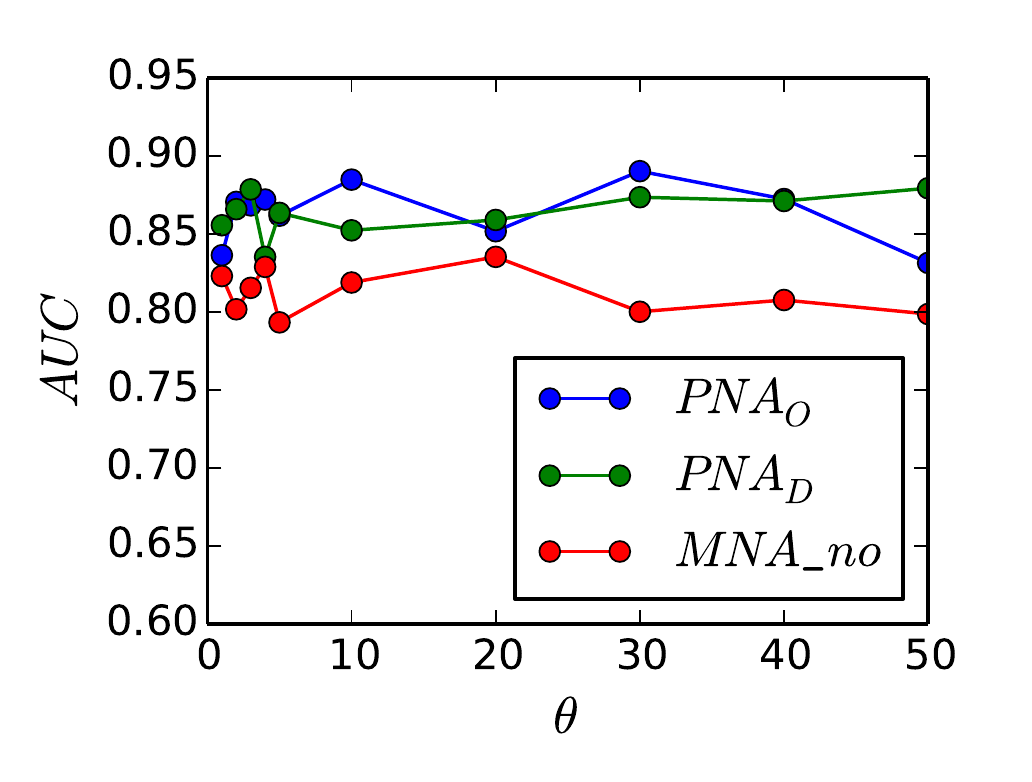}
    \end{minipage}
}
\caption{AUC of different class imbalance link prediction methods.}\label{eg_fig_6}
\end{figure*}

\subsection{Experiment Settings}
In this part, we will talk about the experiment settings in details, which includes: (1) comparison methods, (2) evaluation methods, and (3) experiment setups.
\subsubsection{Comparison Methods}
The comparison methods used in the experiments can be divided into the following $4$ categories:

\noindent \textit{Methods with Generic Stable Matching}:

\begin{itemize}

\item {\ouromt}: {\ouromt} (PNA with \underline{O}ver sampling \& \underline{G}eneric stable \underline{M}atching) is the method proposed in this paper, which consists of two steps: (1) class imbalance link prediction with over sampling, and (2) candidate pruning with \textit{generic stable matching}.

\item {\ourdmt}: {\ourdmt} (PNA with \underline{D}own sampling \& \underline{G}eneric stable \underline{M}atching) is another method proposed in this paper, which consists of two steps: (1) class imbalance link prediction with down sampling, and (2) candidate pruning with \textit{generic stable matching}.

\end{itemize}

\noindent \textit{Methods with Traditional Stable Matching}

\begin{itemize}

\item {\ourom}: {\ourom} (PNA with \underline{O}ver sampling \& traditional stable \underline{M}atching) is identical to {\ouromt} except that in the second step, {\ourom} applies the traditional \textit{stable matching} \cite{GS62, KZY13}.

\item {\ourdm}: {\ourdm} (PNA with \underline{D}own sampling \& traditional stable \underline{M}atching) is identical to {\ourdmt} except that in the second step, {\ourdm} applies the traditional \textit{stable matching} \cite{GS62, KZY13}.

\end{itemize}

\noindent \textit{Class Imbalance Anchor Link Prediction}:

\begin{itemize}

\item {\ouro}: {\ouro} (PNA with \underline{O}ver sampling) is the link prediction method with over sampling to overcome the class imbalance problem and has no matching step.

\item {\ourd}: {\ourd} (PNA with \underline{D}own sampling) is the link prediction method with down sampling to overcome the class imbalance problem and has no matching step.

\end{itemize}

\noindent \textit{Existing Network Anchoring Methods}
\begin{itemize}

\item {\ourm}: {\ourm} (\underline{M}ulti-\underline{N}etwork \underline{A}nchoring) is a \textit{two-phase} method proposed in \cite{KZY13} which includes: (1) supervised link prediction without addressing class imbalance problem; (2) traditional stable matching \cite{GS62, KZY13}.

\item {\oursup}: {\oursup} ({\ourm} without one-to-one constraint) is the first step of {\ourm} proposed in \cite{KZY13} which can predict anchor links without addressing the class imbalance problem and has no matching step.

\end{itemize}

\subsubsection{Evaluation Metrics}
The output of different link prediction methods can be either \textit{predicted labels} or \textit{confidence scores}, which are evaluated by \textit{Accuracy}, \textit{AUC}, \textit{F1} in the experiments.

\subsubsection{Experiment Setups}
In the experiment, initially, a fully aligned network containing $3000$ users in both Twitter and Foursquare is sampled from the datasets. All the existing \textit{anchor links} are grouped into the positive link set and all the possible non-existing \textit{anchor links} are used as the potential link set. Certain number of links are randomly sampled from the potential link set as the negative link set, which is controlled by parameter $\theta$. Parameter $\theta$ represents the $\frac{\#negative}{\#positive}$ rate, where $\theta = 1$ denotes the class balance case, i.e., $\# positive$ equals to $\# negative$; $\theta = 50$ represents that case that negative instance set is $50$ times as large as that of the positive instance set, i.e., $\# negative = 50 \times\# positive$. In the experiment, $\theta$ is chosen from $\{1, 2, 3, 4, 5, 10, 20, 30, 40, 50\}$. Links in the positive and negative link sets are partitioned into two parts with 10-fold cross validation, where $9$ folds are used as the training set and $1$ fold is used as the test set. To simulate the \textit{partial alignment} networks, certain positive links are randomly sampled from the positive training set as the final positive training set under the control of parameter $\eta$. $\eta$ is chosen from $\{0.1, 0.2, \cdots, 1.0\}$, where $\eta = 0.1$ denotes that the networks are $10\%$ aligned and $\eta = 1.0$ shows that the networks are fully aligned. With links in the positive training set, \textit{anchor adjacency tensor} based features and the \textit{latent feature vectors} are extracted from the network to build \textit{link prediction model} $\mathcal{M}$. In building model $\mathcal{M}$, \textit{over sampling} and \textit{under sampling} techniques are applied and the sampling rate is determined by parameter $\sigma \in \{0.0, 0.1, 0.2, \cdots, 1.0\}$, where $\sigma = 0.3$ denotes that $0.3 \times (\#negative - \#positive)$ negative links are randomly removed from the negative link set in under sampling; or $0.3 \times (\#negative - \#positive)$ positive links are generated and added to the positive link set in over sampling. Before applying model $\mathcal{M}$ to the test set, \textit{pre-pruning} process is conducted on the test set in advance. Based on the prediction results of model $\mathcal{M}$ on the test set, \textit{post-pruning} with \textit{generic stable matching} is applied to further prune the non-existent candidates to ensure that the final prediction results across the \textit{partially aligned networks} can meet the $one-to-one_{\le}$ constraint controlled by the \textit{partial matching parameter} $K$.

\begin{figure*}[t]
\centering
\subfigure[Acc.@$\theta = 5$, $\eta = 0.4$]{ \label{eg_fig_7_1}
    \begin{minipage}[l]{0.8\columnwidth}
      \centering
      \includegraphics[width=1.0\textwidth]{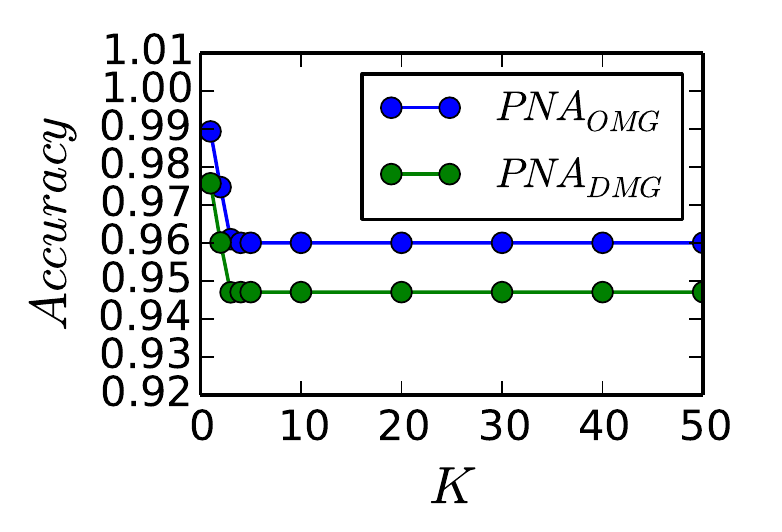}
    \end{minipage}
}
\subfigure[F1@$\theta = 5$, $\eta = 0.4$]{ \label{eg_fig_7_2}
    \begin{minipage}[l]{0.8\columnwidth}
      \centering
      \includegraphics[width=1.0\textwidth]{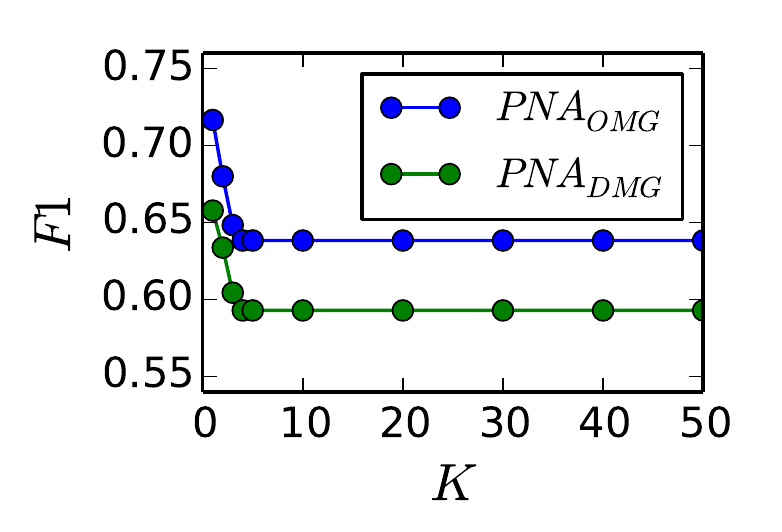}
    \end{minipage}
}
\subfigure[Acc.@$\theta = 50$, $\eta = 0.9$]{ \label{eg_fig_7_3}
    \begin{minipage}[l]{0.8\columnwidth}
      \centering
      \includegraphics[width=1.0\textwidth]{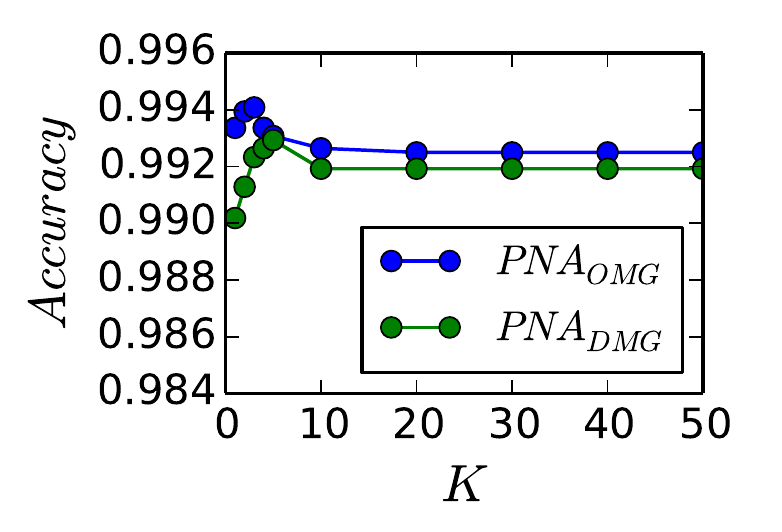}
    \end{minipage}
}
\subfigure[F1@$\theta = 50$, $\eta = 0.9$]{ \label{eg_fig_7_4}
    \begin{minipage}[l]{0.8\columnwidth}
      \centering
      \includegraphics[width=1.0\textwidth]{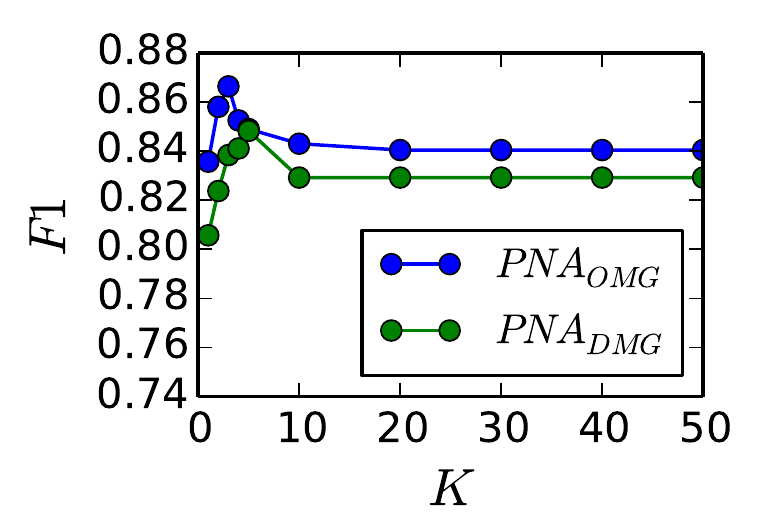}
    \end{minipage}
}
\caption{F1, Accuracy of {\ouromt} and {\ourdmt} with different partial matching rates.}\label{eg_fig_7}
\end{figure*}

\subsection{Experiment Results}
In this part, we will give the experiment results of all these comparison methods in addressing the \textit{partial network alignment} problem. This part includes (1) analysis of sampling methods in class imbalance link prediction; (2) performance comparison of different link prediction methods; and (3) parameter analysis.

\subsubsection{Analysis of Sampling Methods}


To examine whether sampling methods can improve the prediction performance on the imbalanced classification problem or not, we also compare {\ouro}, {\ourd} with {\oursup} and the results are given in Figure~\ref{eg_fig_6}, where we fix $\theta$ as 10 but change $\eta$ with values in $\{0.1, 0.2, \cdots, 1.0\}$ and compare the \textit{AUC} achieved by {\ouro}, {\ourd} and {\oursup}. We can observe that the \textit{AUC} values of all these three methods increases with the increase of $\eta$ but {\ouro} and {\ourd} perform consistently better than {\oursup}. In Figure~\ref{eg_fig_6_2}, we fix $\eta$ as $0.6$ but change $\theta$ with values in \{1, 2, 3, 4, 5, 10, 20, 30, 40, 50\} and compare the \textit{AUC} of {\ouro}, {\ourd} and {\oursup}. As shown in Figure~\ref{eg_fig_6_2}, the performance of {\ouro}, {\ourd} and {\oursup} can all varies slightly with $\theta$ changing from $1$ to $50$ and {\ouro}, {\ourd} can achieve better performance than {\oursup} consistently.




\subsubsection{Comparison of Different Link Prediction Methods}

Meanwhile, as \textit{generic stable matching} based post pruning can only output the labels of potential \textit{anchor links} in the test set, we also evaluate all these methods by comparing their \textit{Accuracy} and \textit{F1} score Tables~\ref{tab:setting1}-\ref{tab:setting2}. In Table~\ref{tab:setting1}, we fix $\theta$ as $10$, $K$ as $5$ but change $\eta$ with values in $\{0.1, 0.2, \cdots, 1.0\}$. Table~\ref{tab:setting1} has two parts. The upper part of Table~\ref{tab:setting1} shows the \textit{Accuracy} achieved by all the methods with various $\eta$, and the lower part shows the \textit{F1} score. Generally, the performance of all comparison methods rises as $\eta$ increases. In the upper part, methods {\ouromt} and {\ourdmt} can consistently perform better than all other comparison methods for different $\eta$. For example, when $\eta = 0.5$, the \textit{Accuracy} achieved by {\ouromt} is higher than {\ourom} by $3.45\%$, higher than {\ourm} by $6.0\%$, higher than {\ouro} by $7.51\%$ and higher than {\oursup} by $7.75\%$; meanwhile, the \textit{Accuracy} achieved by {\ourdmt} is higher than {\ourdm}, {\ourm}, {\ourd} and {\oursup} as well. The advantages of {\ouromt} and {\ourdmt} over other comparison methods are more obvious under the evaluation of \textit{F1} as in \textit{class imbalance} settings, \textit{Accuracy} is no longer an appropriate evaluation metric \cite{C05}. For example, when $\eta = 0.5$, the \textit{F1} achieved by {\ouromt} is about $13.25\%$ higher than {\ourom}, $24\%$ higher than {\ourm}, $101.6\%$ higher than {\ouro} and $165\%$ higher than {\oursup}; so is the case for method {\ourdmt}. The experiment results show that {\ouromt} and {\ourdmt} can work well with datasets containing different ratio of anchor links across the networks. Similar results can be obtained from Table~\ref{tab:setting2}, where we fix $\eta = 0.6$, $K$ as $5$ but change $\theta$ with values in $\{1, 2, 3, 4, 5, 10, 20, 30, 40, 50\}$. It shows that {\ouromt} and {\ourdmt} can effectively address the class imbalance problem.




The fact that (1) {\ouromt} can outperform {\ourom} ({\ourdmt} outperforms {\ourdm}) shows that \textit{generic stable matching} can work well in dealing with \textit{partially aligned social networks}; (2) {\ourom} can beat {\ouro} (and {\ourdm} beats {\ourd}) means that \textit{stable matching} can achieve very good \textit{post-pruning} results; (3) {\ourom} and {\ourdm} can perform better than {\ourm} (or {\ouro} and {\ourd} can achieve better results than {\oursup}) means that sampling methods can overcome the \textit{class imbalance} problem very well.



\begin{table*}[t]

\caption{Performance comparison of different methods for partial network alignment with different network alignment rates.} 
\label{tab:setting1}
\centering
{
\begin{tabular}{lrcccccccccc}
\toprule
\multicolumn{2}{l}{}&\multicolumn{10}{c}{anchor link sampling rate $\eta$}\\
\cmidrule{3-12}
\rotatebox{90}{}&Methods	&0.1	&0.2	& 0.3	& 0.4	&0.5	&0.6	&0.7	&0.8 &0.9 &1.0 \\ 
\midrule
\multirow{8}{*}{\textsc{Acc}}

&{\ouromt}     &\textbf{0.964}     &\textbf{0.966}     &\textbf{0.973}     &\textbf{0.967}     &\textbf{0.987}     &\textbf{0.989}     &\textbf{0.981}     &\textbf{0.988}     &\textbf{0.989}     &\textbf{0.990}     \\
&{\ourdmt}     &\textbf{0.960}     &\textbf{0.974}     &\textbf{0.961}     &\textbf{0.976}     &\textbf{0.983}     &\textbf{0.975}     &\textbf{0.982}     &\textbf{0.989}     &\textbf{0.986}     &\textbf{0.990}     \\
\cmidrule{3-12}
&{\ourom}     &{0.942}     &{0.938}     &{0.948}     &{0.945}     &{0.954}     &{0.960}     &{0.970}     &{0.968}     &{0.983}     &{0.981}     \\
&{\ourdm}     &{0.940}     &{0.951}     &{0.949}     &{0.929}     &{0.949}     &{0.947}     &{0.969}     &{0.966}     &{0.983}     &{0.981}     \\
&{\ourm}     &{0.917}     &{0.918}     &{0.922}     &{0.922}     &{0.931}     &{0.937}     &{0.940}     &{0.943}     &{0.949}     &{0.971}     \\
\cmidrule{3-12}
&{\ouro}     &{0.905}     &{0.907}     &{0.915}     &{0.915}     &{0.918}     &{0.927}     &{0.926}     &{0.925}     &{0.929}     &{0.921}     \\
&{\ourd}     &{0.905}     &{0.908}     &{0.911}     &{0.912}     &{0.915}     &{0.926}     &{0.923}     &{0.925}     &{0.929}     &{0.923}     \\
&{\oursup}     &{0.895}     &{0.899}     &{0.901}     &{0.907}     &{0.916}     &{0.921}     &{0.922}     &{0.924}     &{0.919}     &{0.922}     \\

\midrule
\multirow{8}{*}{F1}

&{\ouromt}     &\textbf{0.280}     &\textbf{0.375}     &\textbf{0.442}     &\textbf{0.496}     &\textbf{0.615}     &\textbf{0.717}     &\textbf{0.776}     &\textbf{0.843}     &\textbf{0.941}     &\textbf{0.965}     \\
&{\ourdmt}     &\textbf{0.283}     &\textbf{0.374}     &\textbf{0.412}     &\textbf{0.481}     &\textbf{0.589}     &\textbf{0.658}     &\textbf{0.783}     &\textbf{0.848}     &\textbf{0.925}     &\textbf{0.972}     \\
\cmidrule{3-12}
&{\ourom}     &{0.230}     &{0.318}     &{0.384}     &{0.452}     &{0.543}     &{0.638}     &{0.723}     &{0.824}     &{0.916}     &{0.963}     \\
&{\ourdm}     &{0.239}     &{0.324}     &{0.369}     &{0.424}     &{0.526}     &{0.593}     &{0.716}     &{0.812}     &{0.919}     &{0.963}     \\
&{\ourm}     &{0.211}     &{0.267}     &{0.375}     &{0.420}     &{0.496}     &{0.578}     &{0.705}     &{0.782}     &{0.899}     &{0.943}     \\
\cmidrule{3-12}
&{\ouro}     &{0.014}     &{0.054}     &{0.211}     &{0.210}     &{0.305}     &{0.402}     &{0.413}     &{0.385}     &{0.428}     &{0.438}     \\
&{\ourd}     &{0.010}     &{0.048}     &{0.131}     &{0.165}     &{0.257}     &{0.380}     &{0.365}     &{0.367}     &{0.405}     &{0.438}     \\
&{\oursup}     &{0.004}     &{0.021}     &{0.042}     &{0.067}     &{0.232}     &{0.322}     &{0.339}     &{0.346}     &{0.360}     &{0.380}     \\

\bottomrule
\end{tabular}
}

\end{table*}

\begin{table*}[t]

\caption{Performance comparison of different methods for partial network alignment with different negative positive rates.} 
\label{tab:setting2}
\centering
{
\begin{tabular}{lrcccccccccc}
\toprule
\multicolumn{2}{l}{}&\multicolumn{10}{c}{negative positive rate $\theta$}\\
\cmidrule{3-12}
Measure&Methods	&1	&2	& 3	&4	&5	&10	&20	&30 &40 &50 \\ 
\midrule
\multirow{8}{*}{\textsc{Acc}}

&{\ouromt}     &\textbf{0.941}     &\textbf{0.900}     &\textbf{0.903}     &\textbf{0.904}     &\textbf{0.905}     &\textbf{0.989}     &\textbf{0.995}     &\textbf{0.995}     &\textbf{0.998}     &\textbf{0.997}     \\
&{\ourdmt}     &\textbf{0.920}     &\textbf{0.917}     &\textbf{0.903}     &\textbf{0.913}     &\textbf{0.893}     &\textbf{0.975}     &\textbf{0.994}     &\textbf{0.998}     &\textbf{0.997}     &\textbf{0.997}     \\
\cmidrule{3-12}
&{\ourom}     &{0.934}     &{0.898}     &{0.899}     &{0.882}     &{0.898}     &{0.960}     &{0.975}     &{0.981}     &{0.992}     &{0.995}     \\
&{\ourdm}     &{0.916}     &{0.914}     &{0.892}     &{0.910}     &{0.887}     &{0.947}     &{0.977}     &{0.981}     &{0.990}     &{0.990}     \\
&{\ourm}     &{0.914}     &{0.863}     &{0.884}     &{0.886}     &{0.878}     &{0.937}     &{0.966}     &{0.970}     &{0.978}     &{0.986}     \\
\cmidrule{3-12}
&{\ouro}     &{0.706}     &{0.795}     &{0.834}     &{0.849}     &{0.880}     &{0.927}     &{0.958}     &{0.970}     &{0.976}     &{0.980}     \\
&{\ourd}     &{0.752}     &{0.812}     &{0.836}     &{0.865}     &{0.875}     &{0.926}     &{0.955}     &{0.968}     &{0.976}     &{0.980}     \\
&{\oursup}     &{0.714}     &{0.781}     &{0.825}     &{0.839}     &{0.873}     &{0.921}     &{0.953}     &{0.968}     &{0.975}     &{0.980}     \\

\midrule
\multirow{8}{*}{F1}

&{\ouromt}     &\textbf{0.943}     &0.870     &\textbf{0.835}     &\textbf{0.805}     &\textbf{0.776}     &\textbf{0.717}     &\textbf{0.608}     &\textbf{0.552}     &\textbf{0.565}     &\textbf{0.524}     \\
&{\ourdmt}     &0.926    &\textbf{0.890}     &\textbf{0.834}     &\textbf{0.821}     &0.754     &\textbf{0.658}     &\textbf{0.602}     &\textbf{0.577}     &\textbf{0.548}     &\textbf{0.533}     \\
\cmidrule{3-12}
&{\ourom}     &{0.936}     &{0.867}     &{0.832}     &{0.772}     &{0.769}     &{0.638}     &{0.550}     &{0.470}     &{0.438}     &{0.366}     \\
&{\ourdm}     &{0.923}     &{0.887}     &{0.822}     &{0.819}     &{0.747}     &{0.593}     &{0.563}     &{0.468}     &{0.419}     &{0.405}     \\
&{\ourm}     &{0.887}     &{0.800}     &{0.790}     &{0.760}     &{0.694}     &{0.578}     &{0.508}     &{0.397}     &{0.346}     &{0.329}     \\
\cmidrule{3-12}
&{\ouro}     &{0.600}     &{0.609}     &{0.553}     &{0.515}     &{0.492}     &{0.402}     &{0.294}     &{0.251}     &{0.131}     &{0.051}     \\
&{\ourd}     &{0.687}     &{0.633}     &{0.569}     &{0.528}     &{0.455}     &{0.380}     &{0.230}     &{0.131}     &{0.093}     &{0.067}     \\
&{\oursup}     &{0.575}     &{0.542}     &{0.526}     &{0.483}     &{0.447}     &{0.322}     &{0.204}     &{0.105}     &{0.075}     &{0.041}     \\

\bottomrule
\end{tabular}
}

\end{table*}

\subsubsection{Analysis of Partial Matching Rate}

In the \textit{generic stable matching}, only top $K$ \textit{anchor link candidates} will be preserved. In this part, we will analyze the effects of parameter $K$ on the performance of {\ouromt} and {\ourdmt}. Figure~\ref{eg_fig_7} gives the results (both \textit{Accuracy} and \textit{F1}) of {\ouromt} and {\ourdmt} by setting parameter $K$ with values in $\{1, 2, 3, 4, 5, 10, 20, 30, 40, 50\}$.

In Figures~\ref{eg_fig_7_1}-\ref{eg_fig_7_2}, parameters $\theta$ and $\eta$ are fixed as $5$ and $0.4$ respectively. From the results, we observe that both {\ouromt} and {\ourdmt} can perform very well when $K$ is small and the best is obtained at $K=1$. It shows that the \textit{anchor link candidates} with the highest confidence predicted by {\ouro} and {\ourd} are the optimal \textit{network alignment results} when $\theta$ and $\eta$ are low. In Figures~\ref{eg_fig_7_3}-\ref{eg_fig_7_4}, we set $\eta$ as $0.9$ and $\theta$ as $50$ (i.e., the networks contain more anchor links and the training/test sets become more imbalance), we find that the performance of both {\ouromt} and {\ourdmt} increases first and then decreases and finally stay stable as $K$ increases, which shows that the optimal \textit{anchor link candidates} are those within the top $K$ candidate set rather than the one with the highest confidence as the training/test sets become more imbalance.

\begin{figure}[t]
\centering
    \begin{minipage}[l]{0.8\columnwidth}
      \centering
      \includegraphics[width=1.0\textwidth]{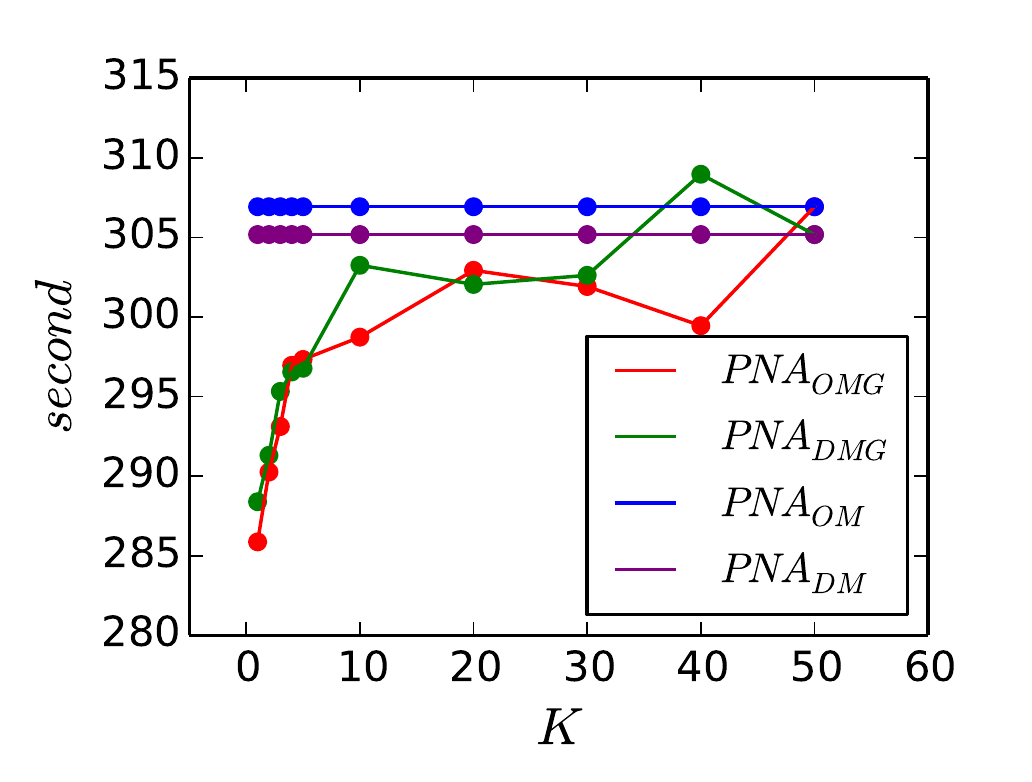}
    \end{minipage}
\caption{Time cost of {\ouromt} and {\ourdmt} with different partial matching rates.}\label{eg_fig_8}
\end{figure}

In addition, the \textit{partial matching strategy} can shrink the preference lists of users a lot, which can lead to lower time cost as shown in Figure~\ref{eg_fig_8} especially for the smaller K values which lead to better accuracy as shown in Figure~\ref{eg_fig_7}.

Results in all these figures show that \textit{generic stable matching} can effectively prune the redundant candidate links and significantly improve the prediction results.

\section{Related Works}
\label{sec:relatedworks}

\textit{Aligned social network} studies have become a hot research topic in recent years. Kong et al. \cite{KZY13} are the first to propose the \textit{anchor link prediction} problem in \textit{fully aligned social networks}. Zhang et al. \cite{ZKY13, ZKY14, ZYZ14, ZY15-3} propose to predict links for new users and new networks by transferring heterogeneous information across \textit{aligned social networks}. A comprehensive survey about link prediction problems across \textit{multiple social networks} is available in \cite{ZY14}. In addition to link prediction problems, Jin and Zhang et al. \cite{JZYYL14, ZY15, ZY15-2} introduce the community detection problems across aligned networks and Zhan et al. \cite{ZZWYX15} study the information diffusion across aligned social networks.



Meta path first proposed by Sun et al. \cite{SBGAH11} has become a powerful tool, which can be applied in either in link prediction problems \cite{SBGAH11, SHAC12} or clustering problems \cite{SHYYW11, SAH12}. Sun et al. \cite{SBGAH11} propose to predict co-author relationship in heterogeneous bibliographic networks based on meta path. Sun et al. extend the link prediction model to relationship prediction model based on meta path in \cite{SHAC12}. Sun et al. \cite{SHYYW11} propose to calculate the similarity scores among users based on meta path in bibliographical network. Sun et al. \cite{SAH12} also apply meta path in clustering problem of heterogeneous information networks with incomplete attributes.


Tensor has been widely used in social networks studies.  Moghaddam et al. \cite{MJE12} propose to apply extended tensor factorization model for personalized prediction of review helpfulness. Liu et al. \cite{LJGM13} present a tensor-based framework for integrating heterogeneous multi-view data in the context of spectral clustering. A more detailed tutorial about tensor decomposition and applications is available in \cite{KB09}.

Class imbalance problems in classification can be very common in real-world applications. Chawla et al. \cite{CBHK02} propose a technique for over-sampling the minority class with generated new synthetic minority instances. Kubat et al. \cite{KM97} propose to address the class imbalance problems with under sampling of the majority cases in the training set. A systematic study of the \textit{class imbalance problem} is available in \cite{JS02}.


College admission problem \cite{R85} and stable marriage problem \cite{GI89} have been studied for many years and lots of works have been done in the last century. In recent years, some new papers have come out in these areas. Sotomayor et al. \cite{S08} propose to analyze the stability of the equilibrium outcomes in the admission games induced by stable matching rules. Ma \cite{M98} analyzes the truncation in stable matching and the small core in nash equilibrium in college admission problems. Flor\'{e}en et al. \cite{FKPS10} propose to study the almost stable matching by truncating the Gale-Shapley algorithm.

\section{Conclusion}\label{sec:conclusion}

In this paper, we study the \textit{partial network alignment} problem across \textit{partially aligned social networks}. 
To address the challenges of the studied problem, a new method {\our} is proposed in this paper. {\our} can extract features for \textit{anchor links} based on a set of \textit{anchor meta paths} and overcome the \textit{class imbalance} problem with \textit{over sampling} and \textit{down sampling}. {\our} can effectively prune the non-existing \textit{anchor links} with \textit{generic stable matching} to ensure the results can meet the $one-to-one_{\le}$ constraint. Extensive experiments done on two real-world \textit{partially aligned networks} show the superior performance of {\our} in addressing the \textit{partial network alignment problem}.


\label{sec:ack}
\section{Acknowledgement}
This work is supported in part by NSF through grants CNS-1115234, Google Research Award, the Pinnacle Lab at Singapore Management University, and Huawei grants.


\begin{thebibliography}{10}

\bibitem{BGGSW09}
M.~Bayati, M.~Gerritsen, D.~Gleich, A.~Saberi, and Y.~Wang.
\newblock Algorithms for large, sparse network alignment problems.
\newblock In {\em ICDM}, 2009.

\bibitem{BG07}
I.~Bhattacharya and L.~Getoor.
\newblock Collective entity resolution in relational data.
\newblock {\em TKDD}, 1(1), 2007.

\bibitem{C05}
N.~Chawla.
\newblock Data mining for imbalanced datasets: An overview.
\newblock In {\em Data Mining and Knowledge Discovery Handbook}. 2005.

\bibitem{CBHK02}
N.~Chawla, K.~Bowyer, L.~Hall, and P.~Kegelmeyer.
\newblock Smote: Synthetic minority over-sampling technique.
\newblock {\em J. Artif. Int. Res.}, 2002.

\bibitem{DF81}
L.~Dubins and D.~Freedman.
\newblock Machiavelli and the gale-shapley algorithm.
\newblock {\em The American Mathematical Monthly}, 1981.

\bibitem{DS13}
M.~Duggan and A.~Smith.
\newblock Social media update 2013.
\newblock 2013.
\newblock Report available at
  \url{http://www.pewinternet.org/2013/12/30/social-media-update-2013/}.

\bibitem{ES07}
J.~Euzenat and P.~Shvaiko.
\newblock {\em Ontology Matching}.
\newblock Springer-Verlag New York, Inc., Secaucus, NJ, USA, 2007.

\bibitem{FKPS10}
P.~Flor\'{e}en, P.~Kaski, V.~Polishchuk, and J.~Suomela.
\newblock Almost stable matchings by truncating the gale-shapley algorithm.
\newblock {\em Algorithmica}, 2010.

\bibitem{GS62}
D.~Gale and L.~Shapley.
\newblock College admissions and the stability of marriage.
\newblock {\em The American Mathematical Monthly}, 1962.

\bibitem{GI89}
D.~Gusfield and R.~Irving.
\newblock {\em The Stable Marriage Problem: Structure and Algorithms}.
\newblock 1989.

\bibitem{JS02}
N.~Japkowicz and S.~Stephen.
\newblock The class imbalance problem: A systematic study.
\newblock {\em Intelligent Data Analysis}, 2002.

\bibitem{JZYYL14}
S.~Jin, J.~Zhang, P.~Yu, S.~Yang, and A.~Li.
\newblock Synergistic partitioning in multiple large scale social networks.
\newblock In {\em IEEE BigData}, 2014.

\bibitem{KB09}
T.~Kolda and B.~Bader.
\newblock Tensor decompositions and applications.
\newblock {\em SIAM REVIEW}, 2009.

\bibitem{KZY13}
X.~Kong, J.~Zhang, and P.~Yu.
\newblock Inferring anchor links across heterogeneous social networks.
\newblock In {\em CIKM}, 2013.

\bibitem{KM97}
M.~Kubat and S.~Matwin.
\newblock Addressing the curse of imbalanced training sets: One-sided
  selection.
\newblock In {\em ICML}, 1997.

\bibitem{LLC10}
R.~Lichtenwalter, J.~Lussier, and N.~Chawla.
\newblock New perspectives and methods in link prediction.
\newblock In {\em KDD}, 2010.

\bibitem{LJGM13}
X.~Liu, S.~Ji, W.~Glanzel, and B.~De Moor.
\newblock Multiview partitioning via tensor methods.
\newblock {\em TKDE}, 2013.

\bibitem{M98}
J.~Ma.
\newblock Stable matchings and the small core in nash equilibrium in the
  college admissions problem.
\newblock Technical report, 1998.

\bibitem{M14}
MarketingCharts.
\newblock Majority of twitter users also use instagram.
\newblock 2014.
\newblock Report available at
  \url{http://www.marketingcharts.com/wp/online/majority-of-twitter-users-also-use-instagram-38941/}.

\bibitem{ME11}
A.~Menon and C.~Elkan.
\newblock Link prediction via matrix factorization.
\newblock In {\em ECML/PKDD}, 2011.

\bibitem{MJE12}
S.~Moghaddam, M.~Jamali, and M.~Ester.
\newblock Etf: Extended tensor factorization model for personalizing prediction
  of review helpfulness.
\newblock In {\em WSDM}, 2012.

\bibitem{PFRE13}
O.~Peled, M.~Fire, L.~Rokach, and Y.~Elovici.
\newblock Entity matching in online social networks.
\newblock In {\em SOCIALCOM}, 2013.

\bibitem{R85}
A.~Roth.
\newblock The college admissions problem is not equivalent to the marriage
  problem.
\newblock {\em Journal of Economic Theory}, 1985.

\bibitem{S08}
M.~Sotomayor.
\newblock The stability of the equilibrium outcomes in the admission games
  induced by stable matching rules.
\newblock {\em International Journal of Game Theory}, 2008.

\bibitem{SAH12}
Y.~Sun, C.~Aggarwal, and J.~Han.
\newblock Relation strength-aware clustering of heterogeneous information
  networks with incomplete attributes.
\newblock {\em VLDB}, 2012.

\bibitem{SBGAH11}
Y.~Sun, R.~Barber, M.~Gupta, C.~Aggarwal, and J.~Han.
\newblock Co-author relationship prediction in heterogeneous bibliographic
  networks.
\newblock In {\em ASONAM}, 2011.

\bibitem{SHAC12}
Y.~Sun, J.~Han, C.~Aggarwal, and N.~Chawla.
\newblock When will it happen?: Relationship prediction in heterogeneous
  information networks.
\newblock In {\em WSDM}, 2012.

\bibitem{SHYYW11}
Y.~Sun, J.~Han, X.~Yan, P.~Yu, and T.~Wu.
\newblock Pathsim: Meta path-based top-k similarity search in heterogeneous
  information networks.
\newblock In {\em VLDB}, 2011.

\bibitem{TGHL13}
J.~Tang, H.~Gao, X.~Hu, and H.~Liu.
\newblock Exploiting homophily effect for trust prediction.
\newblock In {\em WSDM}, 2013.

\bibitem{T76}
I.~Tomek.
\newblock {Two Modifications of CNN}.
\newblock {\em {IEEE Transactions on Systems, Man and Cybernetics}}, 1976.

\bibitem{WS12}
K.~Wilcox and A.~T. Stephen.
\newblock Are close friends the enemy? online social networks, self-esteem, and
  self-control.
\newblock Journal of Consumer Research, 2012.

\bibitem{YTYXL13}
Y.~Yao, H.~Tong, X.~Yan, F.~Xu, and J.~Lu.
\newblock Matri: a multi-aspect and transitive trust inference model.
\newblock In {\em WWW}, 2013.

\bibitem{YCZC13}
J.~Ye, H.~Cheng, Z.~Zhu, and M.~Chen.
\newblock Predicting positive and negative links in signed social networks by
  transfer learning.
\newblock In {\em WWW}, 2013.

\bibitem{ZL13}
R.~Zafarani and H.~Liu.
\newblock Connecting users across social media sites: A behavioral-modeling
  approach.
\newblock In {\em KDD}, 2013.

\bibitem{ZZWYX15}
Q.~Zhan, S.~Wang J.~Zhang, P.~Yu, and J.~Xie.
\newblock Influence maximization across partially aligned heterogenous social
  networks.
\newblock In {\em PAKDD}, 2015.

\bibitem{ZKY13}
J.~Zhang, X.~Kong, and P.~Yu.
\newblock Predicting social links for new users across aligned heterogeneous
  social networks.
\newblock In {\em ICDM}, 2013.

\bibitem{ZKY14}
J.~Zhang, X.~Kong, and P.~Yu.
\newblock Transferring heterogeneous links across location-based social
  networks.
\newblock In {\em WSDM}, 2014.

\bibitem{ZY14}
J.~Zhang and P.~Yu.
\newblock Link prediction across heterogeneous social networks: A survey.
\newblock Technical report, 2014.

\bibitem{ZY15}
J.~Zhang and P.~Yu.
\newblock Community detection for emerging networks.
\newblock In {\em SDM}, 2015.

\bibitem{ZY15-3}
J.~Zhang and P.~Yu.
\newblock Integrated anchor and social link predictions across partially
  aligned social networks.
\newblock In {\em IJCAI}, 2015.

\bibitem{ZY15-2}
J.~Zhang and P.~Yu.
\newblock Mcd: Mutual clustering across multiple heterogeneous networks.
\newblock In {\em IEEE BigData Congress}, 2015.

\bibitem{ZYZ14}
J.~Zhang, P.~Yu, and Z.~Zhou.
\newblock Meta-path based multi-network collective link prediction.
\newblock In {\em KDD}, 2014.

\end{thebibliography}
\end{document}